\documentclass{emulateapj}
\input{epsf}
\submitted{Accepted by ApJ on June 28, 2011}
\shorttitle{Evolutionary tracks of tidally stirred disky dwarf galaxies}

\shortauthors{E. L. {\L}okas et al.}

\begin{document}

\title{Evolutionary tracks of tidally stirred disky dwarf galaxies}

\author{Ewa L. {\L}okas\altaffilmark{1}, Stelios Kazantzidis\altaffilmark{2} and Lucio Mayer\altaffilmark{3}}

\altaffiltext{1}{Nicolaus Copernicus Astronomical Center, 00-716 Warsaw, Poland; lokas@camk.edu.pl}
\altaffiltext{2}{Center for Cosmology and Astro-Particle Physics; and Department of Physics; and Department of Astronomy,
    The Ohio State University, Columbus, OH 43210, USA; stelios@astronomy.ohio-state.edu}
\altaffiltext{3}{Institute for Theoretical Physics, University of Z\"urich, CH-8057 Z\"urich, Switzerland;
        lucio@phys.ethz.ch}

\begin{abstract}
Using collisionless $N$-body simulations, we investigate the tidal
evolution of late-type, rotationally supported dwarfs inside Milky
Way-sized host galaxies. Our study focuses on a wide variety of dwarf
orbital configurations and initial structures. During the evolution,
the disky dwarfs undergo strong mass loss, the stellar disks are
transformed into spheroids, and rotation is replaced by random motions
of the stars. Thus, the late-type progenitors are transformed into
early-type dwarfs as envisioned by the tidal stirring model for the
formation of dwarf spheroidal (dSph) galaxies in the Local Group.
We determine the photometric properties of the dwarfs, including the
total visual magnitude, the half-light radius and the central surface
brightness as they would be measured by an observer near the galactic
center. Special emphasis is also placed on studying their kinematics
and shapes. We demonstrate that the measured values are biased by a
number of observational effects including the increasing angle of the
observation cone near the orbital pericenter, the fact that away from
the pericenter the tidal tails are typically oriented along the
line of sight, and the fact that for most of the evolution the stellar
components of the dwarfs are triaxial ellipsoids whose major axis
tumbles with respect to the line of sight. Finally, we compare the
measured properties of the simulated dwarfs to those of dwarf galaxies
in the Local Group. The evolutionary tracks of the dwarfs in different
parameter planes and the correlations between their different
properties, especially the total magnitude and the surface brightness,
strongly suggest that present-day dSph galaxies may have indeed formed
from late-type progenitors as proposed by the tidal stirring scenario.

\end{abstract}


\section{Introduction}

The population of dwarf galaxies of the Local Group (for a review see Mateo 1998; Tolstoy et al. 2009)
can be divided into two rather distinct subgroups, the dwarf irregular (dIrr) galaxies occupying
larger distances from the big
galaxies, the Milky Way and Andromeda, and dwarf spheroidal (dSph) objects clustering near their respective
hosts. While the former are typically disky, bright, dominated by rotation, gas-rich and still forming stars,
the latter are rounder, faint, supported mainly by random motions, gas-poor and dominated by old stellar populations.

The dIrr galaxies are believed to be of primordial origin, i.e. formed by gas accretion onto potential wells of
small dark matter halos. The formation of dSph galaxies, on the other hand, probably required
some evolution that may occur either
in isolation by purely baryonic processes such as cooling, star formation, feedback from supernovae
and UV background radiation (Ricotti \& Gnedin 2005; Tassis et al. 2008; Sawala et al. 2010)
or be induced by gravitational interactions with the environment
in the form of galaxy harassment (e.g., Moore et al. 1996)
resonant stripping (D'Onghia et al. 2009) or tidal stirring (Mayer et al. 2001a,b; Kazantzidis et al. 2004).

The tidal stirring scenario proposed by Mayer et al. (2001a,b) is particularly promising as it provides a link
between the two types of dwarf galaxies observed in the Local Group. It postulates that dIrrs were actually
progenitors of dSph galaxies and provides a mechanism for such transformation in the form of tidal stripping and
tidal shocking of the dwarfs on orbits around the hosts. As the efficiency of these mechanisms increases with
the decreasing distance of the dwarf galaxy from its host, it also naturally explains the morphology-distance relation
observed in the Local Group. When combined with ram pressure stripping and the effect of the UV
background (Bullock et al. 2000; Susa \& Umemura 2004) the model can also account for the low gas content
and the very high mass-to-light ratios of some dwarfs (Mayer et al. 2006, 2007; see also Mayer 2010 for a review).

The efficiency of the tidal transformation of disky dwarfs into dSph galaxies has been recently explored in great
detail by Kazantzidis et al. (2011, hereafter K11) for different orbital and structural configurations
of the dwarfs interacting with the Milky Way. The dwarfs were initially composed of a stellar
disk embedded in a more extended dark matter halo. They confirmed the general picture sketched earlier by
Mayer et al. (2001a,b) and Klimentowski et al. (2009a): in most cases the tidal interaction of the dwarf
with the Milky Way results in strong mass loss, the morphological transformation of the disk to a bar and then
a spheroid and the transition from ordered (rotation) to random motion of the stars. Only in three out of 19
simulated cases (where the orbits were very extended or the dwarf's halo very concentrated)
did the end product retain enough of its initial disky characteristics that it could not be
classified as a dSph galaxy. Similarly, in almost all cases, the morphological transformation involved the
formation at the first pericenter passage of a tidally induced bar that was shortened in the subsequent
evolution.

The study of K11 focused on the intrinsic properties of the dwarfs which are best suited to describe their
dynamics as precisely as possible, but can only be measured accurately in the idealized situation of $N$-body
simulations where full 3D information is available. Here, we focus on strictly observational parameters that can
be directly compared to real data. The purpose of the present work is therefore two-fold. First, we will
measure the observational parameters of the dwarfs in a realistic way
and identify possible biases affecting the measurements in the environment characteristic for dwarf galaxies
in the Local Group. Second, we will explore the correlations between the observed parameters of the dwarfs in
order to see if these can provide some insight into their formation scenarios.

The paper is organized as follows. In section 2 we briefly describe the simulations used in this work. Section
3 discusses the way we measure the observational parameters focusing on different observational biases that
affect them. We also describe their variability on different timescales and its origin. In section 4 we present
long-term evolution of the measured parameters for different orbits and different initial structure of the dwarfs.
In section 5 we compare our results to observations. The discussion follows in the last section.

\begin{figure}
\begin{center}
    \leavevmode
    \epsfxsize=7.1cm
    \epsfbox[10 20 220 640]{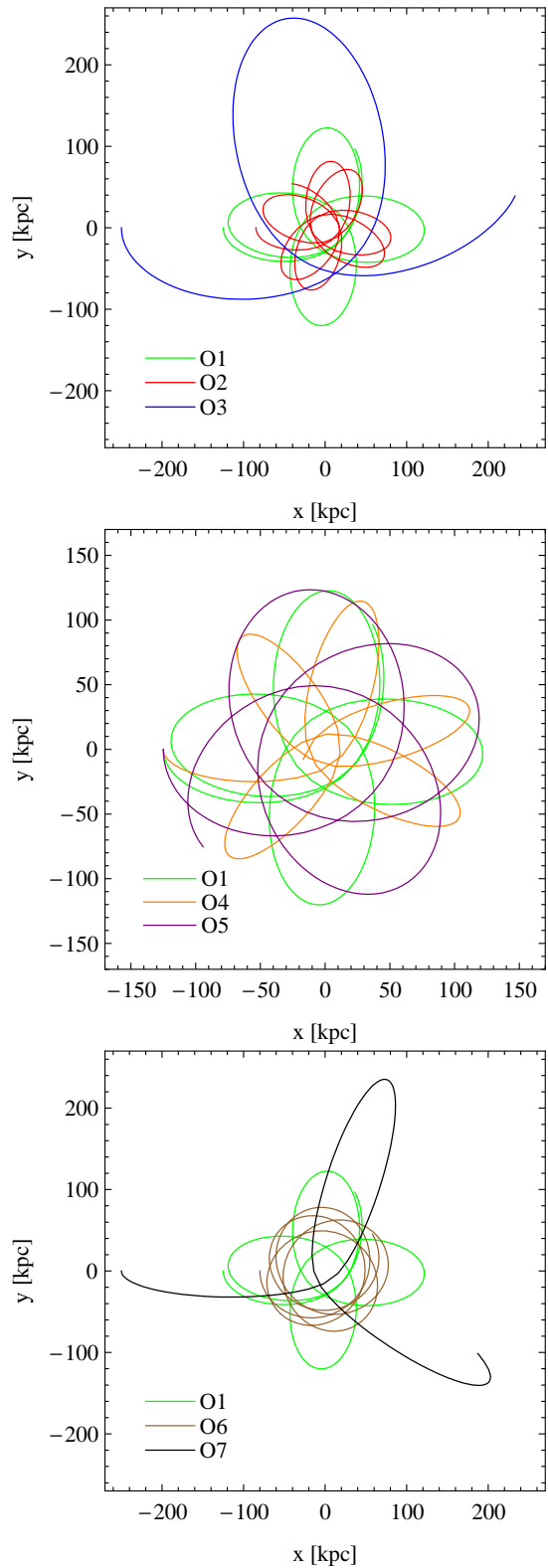}
\end{center}
\caption{Orbits of the dwarf galaxy in simulations O1-O7. In all panels we repeat the orbit O1 for reference.
In the upper panel all three orbits have the same eccentricity $r_{\rm apo}/r_{\rm peri} = 5$
(typical for Milky Way subhalos) but different size. In the
middle panel all three orbits have apocenter $r_{\rm apo} = 125$ kpc but different eccentricity. The lower panel shows
the least (O6) and the most (O7) eccentric orbit in comparison to the typical one (O1). Note that the scale of the
middle panel is different than in of the other two.}
\label{orbits1-7}
\end{figure}

\begin{table*}
\begin{center}
\caption{Properties of the simulated dwarfs. }
\begin{tabular}{llccccccccccl}
\hline
\hline
Simulation & Varied    & $r_{\rm apo}$ & $r_{\rm peri}$ & $T_{\rm orb}$ & $t_{\rm la}$ & $r_{\rm lim}$
& $M_V$ & $r_{1/2}$ & $\mu_V$ & $V/\sigma$ & $e=1-b/a$ &  Color\\
 & parameter & [kpc]       & [kpc]       & [Gyr]         & [Gyr]        & [kpc]         & [mag] &  [kpc]    &
[mag arcsec$^{-2}$]  &  &  & \\
\hline
O1  & orbit                       &   125 &    25  & 2.09 & \ \ 8.35 & 6.28 &$ -11.7$ & 0.36 & 23.4 & 0.36 & 0.20 & green   \\
O2  & orbit                       &\ \ 87 &    17  & 1.28 & \ \ 8.95 & 6.28 &$ -10.2$ & 0.33 & 24.6 & 0.08 & 0.03 & red     \\
O3  & orbit                       &   250 &    50  & 5.40 & \ \ 5.40 & 6.28 &$ -12.8$ & 0.44 & 22.5 & 1.30 & 0.66 & blue    \\
O4  & orbit                       &   125 &\ \ 12.5& 1.81 & \ \ 9.05 & 6.28 &$ -10.4$ & 0.60 & 24.7 & 0.50 & 0.05 & orange  \\
O5  & orbit                       &   125 &    50  & 2.50 &    10.00 & 6.28 &$ -12.3$ & 0.41 & 22.9 & 0.81 & 0.55 & purple  \\
O6  & orbit                       &\ \ 80 &    50  & 1.70 & \ \ 8.50 & 6.28 &$ -12.3$ & 0.47 & 23.3 & 0.37 & 0.35 & brown   \\
O7  & orbit                       &   250 &\ \ 12.5& 4.55 & \ \ 9.10 & 6.28 &$ -11.8$ & 0.46 & 23.6 & 0.62 & 0.39 & black   \\
S6  & $i(-45^{\circ})$            &   125 &    25  & 2.09 & \ \ 8.35 & 6.28 &$ -11.8$ & 0.30 & 22.7 & 0.18 & 0.20 & green   \\
S7  & $i(+45^{\circ})$            &   125 &    25  & 2.09 & \ \ 8.35 & 6.28 &$ -12.0$ & 0.45 & 23.3 & 0.25 & 0.26 & red     \\
S8  & $z_{\rm d}/R_{\rm d}(-0.1)$ &   125 &    25  & 2.09 & \ \ 8.35 & 6.28 &$ -11.8$ & 0.34 & 23.2 & 0.62 & 0.27 & blue    \\
S9  & $z_{\rm d}/R_{\rm d}(+0.1)$ &   125 &    25  & 2.09 & \ \ 8.35 & 6.28 &$ -11.7$ & 0.38 & 23.7 & 0.55 & 0.17 & orange  \\
S10 & $m_{\rm d}(-0.01)$           &   125 &    25  & 2.09 & \ \ 8.35 & 6.28 &$ -10.9$ & 0.38 & 24.5 & 0.45 & 0.09 & purple  \\
S11 & $m_{\rm d}(+0.02)$           &   125 &    25  & 2.09 & \ \ 8.35 & 6.28 &$ -12.8$ & 0.37 & 22.4 & 0.71 & 0.42 & magenta \\
S12 & $\lambda(-0.016)$           &   125 &    25  & 2.09 & \ \ 8.35 & 3.78 &$ -12.3$ & 0.22 & 21.8 & 0.64 & 0.25 & cyan    \\
S13 & $\lambda(+0.026)$           &   125 &    25  & 2.09 & \ \ 8.35 & 6.94 &$ -11.1$ & 0.50 & 24.8 & 0.26 & 0.12 & pink    \\
S14 & $c(-10)$                    &   125 &    25  & 2.10 & \ \ 8.40 & 6.28 &$ -11.4$ & 0.35 & 23.7 & 0.31 & 0.14 & black   \\
S15 & $c(+20)$                    &   125 &    25  & 2.08 & \ \ 8.30 & 6.28 &$ -12.2$ & 0.37 & 23.2 & 0.96 & 0.32 & gray    \\
S16 & $M_{\rm h}(\times 0.2)$     &   125 &    25  & 2.14 & \ \ 8.55 & 3.67 &$ -10.1$ & 0.25 & 24.4 & 0.37 & 0.17 & brown   \\
S17 & $M_{\rm h}(\times 5)$       &   125 &    25  & 1.88 & \ \ 9.40 & 7.00 &$ -13.1$ & 0.48 & 22.8 & 0.63 & 0.18 & yellow  \\
\hline
\label{properties}
\end{tabular}
\end{center}
\end{table*}
% !!!!!!!!!!!!!!!     rows for S12-13 were changed to S14-15 and vice versa to conform with the new notation

\section{The simulations}

In this study, we use a set of the collisionless $N$-body
simulations presented in detail by K11. The
goal of the K11 study was to elucidate the formation of dSph
 galaxies via tidal interactions between late-type,
rotationally supported dwarfs and Milky Way-sized hosts.

K11 employed the method of Widrow \& Dubinski (2005) to construct
numerical realizations of fully self-consistent dwarf galaxy models
composed of exponential stellar disks embedded in cuspy,
cosmologically motivated Navarro et al. (1996, hereafter NFW) dark
matter halos. The reference dwarf galaxy model had a virial mass
of $M_{\rm h} =10^9$ M$_{\odot}$ and a concentration parameter $c=20$.
The disk mass fraction, $m_{\rm d}$, and the halo spin
parameter, $\lambda$, were equal to $0.02$ and $0.04$,
respectively. The resulting disk radial scale length was $R_{\rm d} =
0.41$ kpc (Mo et al. 1998) and the disk thickness was specified by the
thickness parameter $z_{\rm d}/R_{\rm d} = 0.2$, where $z_{\rm d}$ denotes the
(sech$^2$) vertical scale height of the disk. We refer the reader to
K11 for a detailed discussion regarding these choices.

This default dwarf galaxy model was placed on seven different orbits
around a single, massive host with the present-day structural
properties of the Milky Way. In particular, the primary galaxy model was
constructed as a live realization of the MWb model in Widrow \&
Dubinski (2005), which satisfies a broad range of observational
constraints for the Galaxy and consists of an exponential stellar
disk, a Hernquist (1990) bulge, and an NFW dark matter halo.  The
orbital apocenters, $r_{\rm apo}$, and pericenters, $r_{\rm peri}$, of
the dwarf around the host galaxy are listed in the third and fourth
column of Table~\ref{properties}, and their choices are discussed
thoroughly in K11.  The trajectories of the reference dwarf galaxy
model on the seven orbits are shown in Figure~\ref{orbits1-7}
projected onto the initial orbital plane.

The simulations marked as O1-O5 in Table~\ref{properties} correspond to runs R1-R5 in K11. Runs
O6-O7 are two additional experiments mentioned in section 5.4 of K11
and discussed in more detail in {\L}okas et al. (2010c). Lastly, the
alignments of the internal angular momentum of the dwarf, that of the
primary disk and the orbital angular momentum were all mildly prograde
and equal to $i=45^{\circ}$ (see K11 for a discussion regarding the
implications of this choice). Therefore, the results of the present
study should not be affected by any strong coupling of angular
momenta. Initial conditions for all numerical simulations were
generated by building models of dwarf galaxies and placing them at the
apocenters of their orbits. The tidal evolution of the
dwarfs inside their host galaxy was followed for $10$~Gyr using the
multistepping, parallel, tree $N$-body code PKDGRAV (Stadel 2001).

In simulations S6-S17 (see Table~\ref{properties}), we varied the
structural parameters of the dwarfs while keeping them on the same
orbit (as in run O1 with $r_{\rm apo} = 125$ kpc and $r_{\rm peri} =
25$ kpc). These experiments correspond to runs R6-R17 in K11. The
second column of Table~\ref{properties} lists the parameters which
were varied in each case; in parentheses we give the value by which a
given parameter was changed.  Columns 5 and 6 of the Table list the
orbital times and the time of the last apocenter for each simulation.

Lastly, each dwarf galaxy model contained a total of $2.2$ million
particles ($N_{\rm h} = 10^6$ dark matter particles and $N_{\rm d} = 1.2 \times 10^6$ disk
particles). The gravitational softening was set to $\epsilon_{\rm h}=60$~pc
and $\epsilon_{\rm d}=15$~pc for the particles in the two components,
respectively. In addition, the simulations analyzed here use $N_{\rm D}=10^{6}$ particles
in the disk, $N_{\rm B}=5\times10^{5}$ in the bulge, and $N_{\rm H}=2\times10^{6}$
in the dark matter halo of the host galaxy MWb, and employ a gravitational
softening of $\epsilon_{\rm D}=50$~pc, $\epsilon_{\rm B}=50$~pc, and
$\epsilon_{\rm H}=2$~kpc, respectively. The choice for the fairly large
softening in the dark matter particles of the primary galaxy was motivated by
our desire to minimize discretness noise in the host potential.

\begin{figure}
\begin{center}
    \leavevmode
    \epsfxsize=7.5cm
    \epsfbox[60 10 310 725]{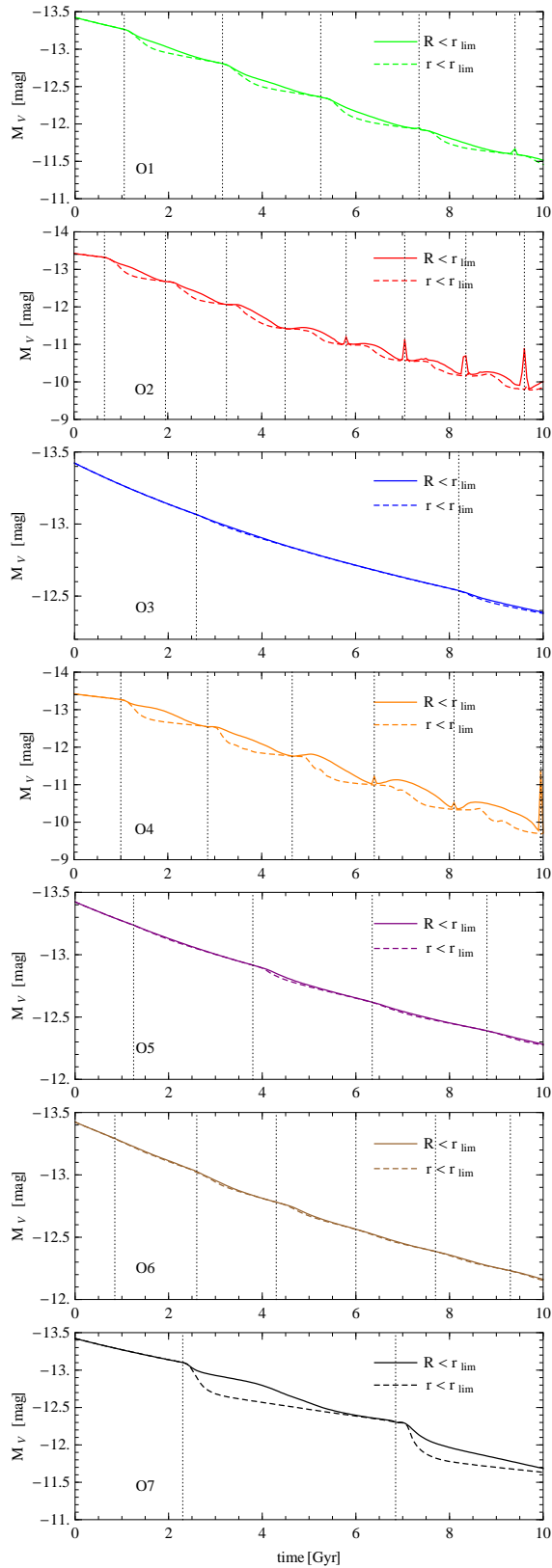}
\end{center}
\caption{The evolution of the total visual magnitude in time for simulations O1-O7.
The dashed line shows measurements inside a sphere
of radius $r_{\rm lim}$ centered on the dwarf where $r_{\rm lim}$ is the limiting radius containing all stars at
the beginning of the simulation. The solid line shows measurements for an observer situated at the galactic center
within the angular aperture corresponding to the projected radius $R = r_{\rm lim}$.
Vertical dotted lines indicate pericenter passages.}
\label{magnitudes1-7}
\end{figure}

\section{Observing the dwarfs}

\subsection{The absolute magnitude}

In order to estimate the observational parameters of the dwarfs we place an imaginary observer at the center of the
Milky Way and introduce a spherical coordinate system centered on this point, with the radial coordinate along the line
connecting the center of the Milky Way with the dwarf and the angles corresponding to celestial coordinates
usually applied in astronomy. Since the dwarfs typically lose a substantial fraction of their stars the measurement of
the total magnitude of the dwarf galaxy as a function of time is non-trivial as it is not a priori obvious which
stars should be included as members of the dwarf at a given instant of the evolution. In real observations this
measurement is done by integrating light up to a given radius or up to infinity if an analytical approximation of the
light distribution is used. Here we adopt a simple approach of counting all stars included within some fixed
limiting radius $r_{\rm lim}$. We adopt the value of this radius to be the distance from the center of the dwarf of the
most distant star in the disk at the starting point of the evolution. The values of this radius are listed in the
seventh column of Table~\ref{properties}. This choice is physically well justified: all the stars that are beyond this
radius at any later stage were ejected due to tidal forces.

Examples of the evolution of the total magnitude of the dwarf on orbits O1-O7 measured within radii $r < r_{\rm
lim}$ from the center of the dwarf as a function of time are illustrated by dashed lines in Figure~\ref{magnitudes1-7}.
The values were obtained by counting the stars within $r_{\rm lim}$, multiplying by the stellar mass and converting to
the luminosity assuming a variable mass-to-light ratio for stars of $M/L_V = (1 + 0.15 t) $M$_{\odot}$/L$_{\odot}$
in the visual band, where $t$ is the time from the start of the simulation, so that $M/L_V = 1 $M$_{\odot}$/L$_{\odot}$
at the start of the simulation (10 Gyr ago) and $M/L_V = 2.5 $M$_{\odot}$/L$_{\odot}$ at the end of the simulation
(which we assume to correspond to the present time). This choice approximates well the evolution
of the simple low-metallicity stellar population in the standard model described by Bruzual \& Charlot (2003, their
Figure 1). In this way we take into account the fading of the stellar population in time.

The evolution of the total magnitude determined in this way is very regular. Until the first pericenter passage
in all cases the
number of stars (and the stellar mass) remains constant and equal to the initial value (almost no stars are stripped
beyond $r_{\rm lim}$) so the decreasing magnitudes are due entirely to the fading of the stellar population.
In the runs on more eccentric orbits (O1, O2, O4 and O7) soon after each pericenter passage (marked by vertical dotted
lines in Figure~\ref{magnitudes1-7}) the luminosity decreases due to the ejection of stars beyond $r_{\rm lim}$
by tidal forces, then declines more slowly again until the next pericenter. For the other runs this dependence on the
orbital position would only be seen if we adopted a constant $M/L$ for the stars while in Figure~\ref{magnitudes1-7}
it is dominated by the fading of the stellar population.
Note that the vertical scale is different in each panel of the Figure, e.g. many
more stars are lost on the tightest orbit (O2) and the one with the smallest pericenter (O4) compared to the
reference orbit O1 while very few are ejected on the most extended orbit O3.

An observer near the galactic center will obviously not be able to measure this intrinsic luminosity of the dwarfs but
rather will count all the stars within the observation cone with the opening angle corresponding to $r_{\rm lim}$ at
the distance of the dwarf. We assume that the distance of the dwarf is known perfectly (without any error)
and the observer counts
all stars within the projected radius $R < r_{\rm lim}$ corresponding to the angular distance from the center of the
dwarf in the plane of the sky, which initially contained all the stars. The corresponding results for orbits O1-O7 are
shown in Figure~\ref{magnitudes1-7} as solid lines. For runs O3, O5 and O6, where the fading dominates, the result
is almost identical, but for the other cases this measurement is biased with respect to the
intrinsic one described above. Both measurements coincide at the pericenters of the orbits, but in other parts of the
orbit (except for the initial part of the evolution) the estimated magnitudes are brighter for the realistic
observations. This is due to the fact that near pericenters the tidal tails are oriented along the orbit, i.e.
perpendicular to the line of sight of the observer, while near apocenters their orientation is more along the
observer's line of sight. This effect, described in more detail by Klimentowski et al. (2009b), is significantly
stronger for orbits with higher eccentricity.

The second observational effect is the presence of the spikes, sudden increases of the measured magnitude at the later
pericenter passages, well visible especially for O2 (second panel of Figure~\ref{magnitudes1-7}). These are the result
of the particularly strong stellar mass loss for this tight orbit combined with a large opening angle of the
observation cone at pericenters (the distance of the dwarf from the galactic center is then about 17 kpc while its
assumed diameter is 12.5 kpc). At later times there is a lot of debris around the Milky Way stripped from the dwarf at
earlier passages (see the orbit plotted with the red line in the upper panel of Figure~\ref{orbits1-7}) which are
included in the observer's field of view. Note that we include here all the debris stars without discriminating by
distance (since distances are typically not known for individual stars) and most often the older debris lies quite far
from the dwarf itself. These stars, if far enough, could in principle be excluded from the sample by careful analysis
of the color-magnitude diagrams of the real dwarfs as they would cause a spread in luminosity. This effect would however
have to be disentangled from the effects of star formation history (see e.g. the discussion for Leo I in Gallart et al.
1999, Sohn et al. 2007 and Mateo et al. 2008).

Given that galaxies on eccentric orbits spend a small
fraction of the orbital time at pericenters and such small pericenter distances are probably not common among Milky Way
satellites (e.g. Lux et al. 2010) this effect should be rare. It could in principle
be observed e.g. for the Sagittarius dwarf if it had completed many pericentric passages around the Milky Way,
which is unlikely (see {\L}okas et al. 2010b). In the following we will use the measured values
at apocenters, where dwarfs are most likely to be seen, as most representative for the observed magnitudes. At these
parts of the orbits the differences between the values measured by our observer ($R < r_{\rm lim}$) and the intrinsic
ones ($r < r_{\rm lim}$) are never larger than 0.4 mag (for run O4) so the tidal tails do not seem to bias the
measurements very strongly even for the most heavily stripped dwarfs.

\begin{figure}
\begin{center}
    \leavevmode
    \epsfxsize=7.4cm
    \epsfbox[60 10 311 737]{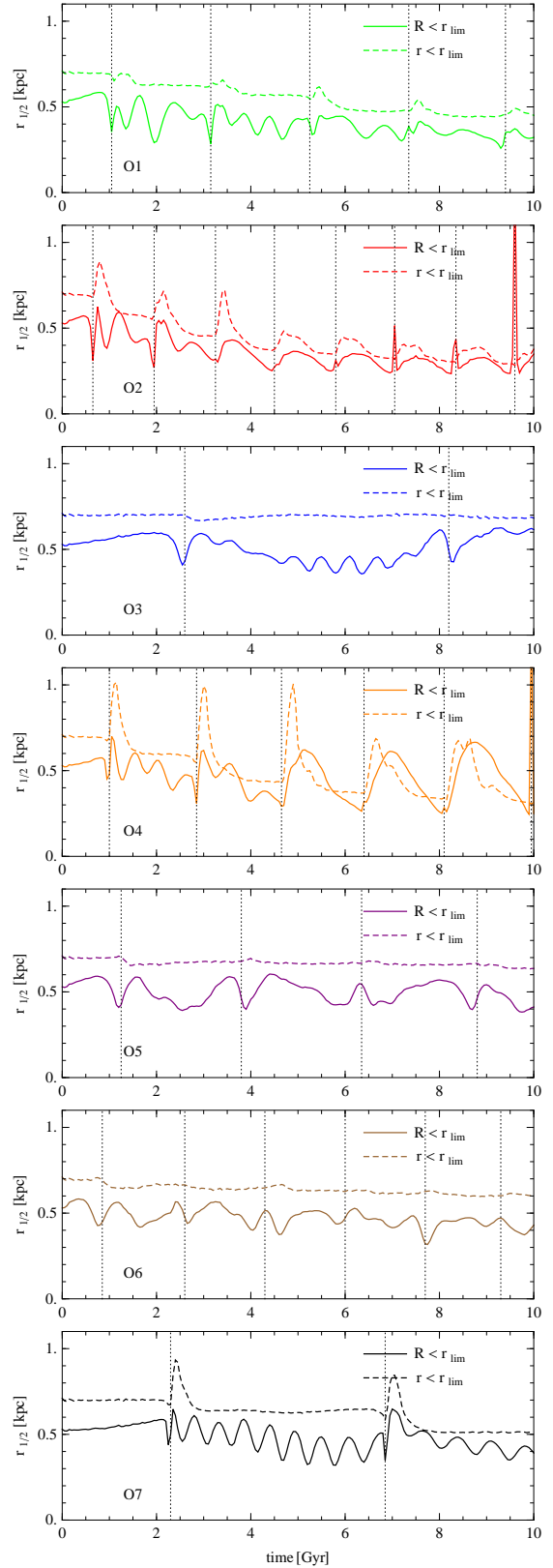}
\end{center}
\caption{The evolution of the half-light radius in time for simulations O1-O7.
The dashed line shows measurements inside a sphere
of radius $r_{\rm lim}$ centered on the dwarf where $r_{\rm lim}$ is the limiting radius containing all stars at
the beginning of the simulation. The solid line shows measurements for an observer situated at the galactic center
within the angular aperture corresponding to the projected radius $R = r_{\rm lim}$. Vertical dotted lines indicate
pericenter passages.}
\label{radii1-7}
\end{figure}

\subsection{The half-light radius}

The next structural parameter of special interest is the half-light radius $r_{1/2}$. We first discuss the half-light
radius measured in 3D, i.e. the radius containing half the total luminosity calculated from the number of stars
contained within the 3D fixed radius $r < r_{\rm lim}$. Half-light radii determined in this way are plotted as dashed
lines in Figure~\ref{radii1-7} for orbits O1-O7. The general trend, best seen for the tightest orbit O2, is that
$r_{1/2}$ decreases with time on a large timescale, which reflects the decreasing size of the dwarf due to tidal
stripping. Significant temporary increase of $r_{1/2}$ is however visible after each pericenter passage (indicated by
vertical dotted lines in Figure~\ref{radii1-7}), especially for runs O4, O7 and O2 with the smallest pericenters.
This is due to the expansion of the dwarf's stellar component in response to the tidal forces which are strongest
at the pericenter.

A realistic measurement of $r_{1/2}$, by an observer located at the center of the Milky Way, would be done in 2D, on
the surface of the sky and using the total luminosity determined within $R < r_{\rm lim}$. These measurements are shown
in Figure~\ref{radii1-7} as solid lines. The half-light radii determined in this way are typically lower that their 3D
counterparts which is due to the fact that the projected density profile of the stars is steeper, i.e. in the center of
the dwarf all stars along the line of sight contribute and not only those that are close to the center in 3D. At later
pericenters of simulation O2 the values of the projected $r_{1/2}$ show strong peaks. This is a direct consequence of
the peaks in the measured total magnitudes due to large opening angles of the observation cone at these instances, as
discussed above. Another feature of the evolution of the projected $r_{1/2}$ are wider bumps between pericenters,
especially well visible in run O4 (but also present in O2), where the measured values of projected $r_{1/2}$ are
about twice larger at apocenter than at pericenter. These are due to increased number of stars from the tidal tails that
for this highly eccentric orbit are oriented along the observer's line of sight for most of the time.

Another interesting observational effect that manifests itself in the measured values of the 2D half-light radii is
their strong variability or even oscillation at the early stages of the evolution, especially visible after the first
and second pericenter. As discussed in detail by K11, in most simulations studied here after the
first pericenter the stellar disk transforms into a triaxial shape, usually a prolate spheroid or a bar which, as the
evolution proceeds, becomes more spherical. The tumbling of the ellipsoid causes the observed variation of the measured
$r_{1/2}$: when the observer's line of sight is perpendicular to the major axis of the stellar component the dwarf
appears more extended and the measured half-light radius is larger; when the major axis is aligned with the observer's
line of sight, the measured $r_{1/2}$ is smaller. This effect is strongest for simulation O7, but present even
in the case of run O3 where a bar does not form and the stellar component is oblate, but still triaxial.
Note that in the case of simulations O2 and O4, where
the dwarf is particularly strongly stirred and its stellar component becomes almost spherical after the third
pericenter, this variability of $r_{1/2}$ is no longer seen at these later stages and the variation of $r_{1/2}$
follows that of the total magnitude.

\begin{figure}
\begin{center}
    \leavevmode
    \epsfxsize=8.2cm
    \epsfbox[60 0 310 505]{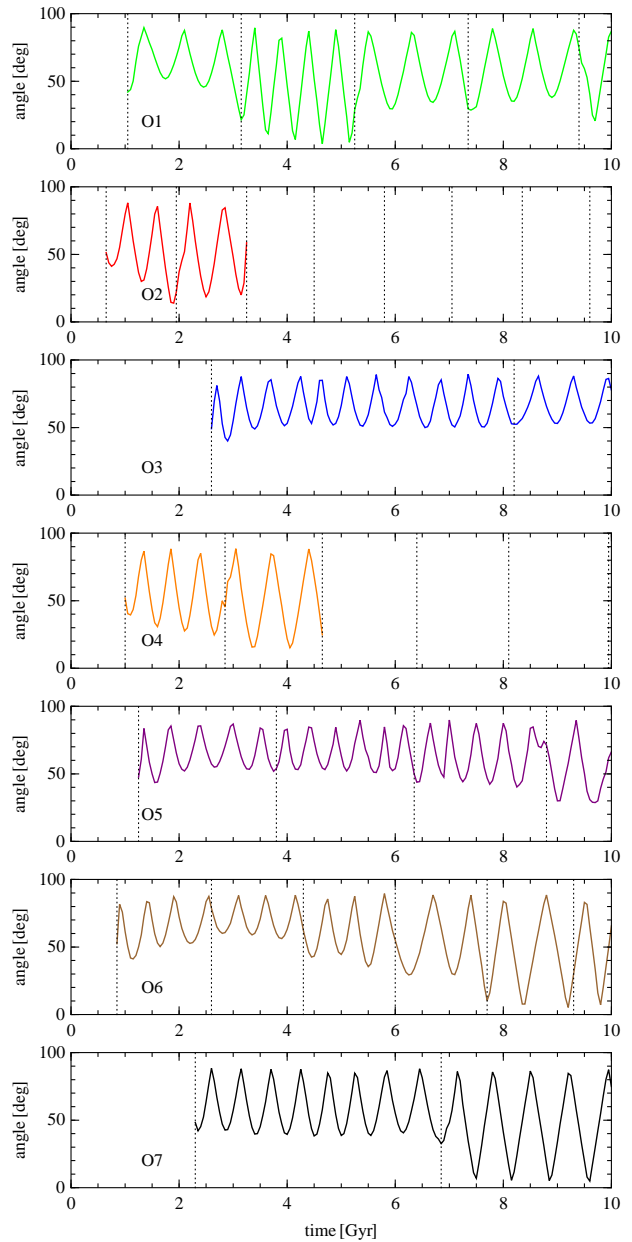}
\end{center}
\caption{The evolution of the angle between the major axis of the stellar component of the dwarf
and the fixed $x$ axis of the
simulation box in time for simulations O1-O7. Measurements are performed for stars within a fixed radius $r < 0.6$
kpc. Vertical dotted lines indicate pericenter passages.}
\label{ba1-7}
\end{figure}

This interpretation of the measurements is confirmed by Figure~\ref{ba1-7} where
we plot the angle between the major axis of the stellar component and the fixed $x$ axis of the simulation box
which lies in the orbital plane of all simulations (at the initial configuration).
The direction of the major axis was determined
using all stars within a fixed radius of 0.6 kpc which corresponds to $(1-2) r_{1/2}$ in all cases. At all times we
measured the angle between the $x$ axis and the dwarf's major axis on the side of the dwarf which is closer to the $x$
axis so all angle values are between $0^{\circ}$ and $90^{\circ}$. The values of the angle were plotted starting
from the first pericenter when the shape of the dwarf's stellar component changes from the disk to the triaxial
ellipsoid so that the major axis is well defined. For orbits O2 and O4 we do not show the measurements until the
end because on these tight orbits the dwarf becomes almost spherical early on and the major axis is then no longer well
defined again. The amplitude of the angle changes in some cases, which is due
to the precession of the rotation axis. In addition, the rate at which the angle changes increases after the second
pericenter for O1 which is due to speeding up of the figure rotation of the ellipsoid in this case, caused by
the tidal force, as discussed by K11 (their section 5.2).

\subsection{The central surface brightness}

\begin{figure}
\begin{center}
    \leavevmode
    \epsfxsize=8.2cm
    \epsfbox[60 0 310 505]{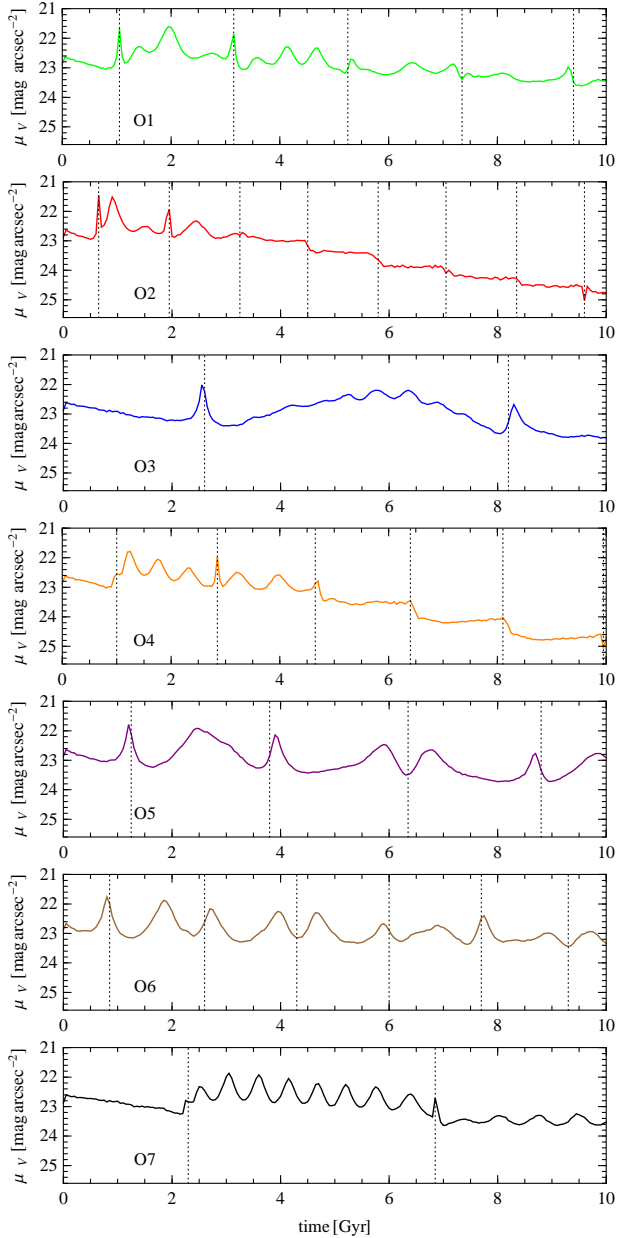}
\end{center}
\caption{The evolution of the central surface brightness in time for simulations O1-O7.
Measurements are performed within $0.1 r_{1/2}$ where $r_{1/2}$ is determined by
an observer situated at the galactic center
within the angular aperture corresponding to the projected radius $R = r_{\rm lim}$.
Vertical dotted lines indicate pericenter passages.}
\label{sb1-7}
\end{figure}

Another characteristic photometric property of dwarf galaxies is the central surface brightness $\mu_V$. This
parameter does not have any 3D analog so we measure it directly in 2D by counting stars within $0.1 r_{1/2}$ (where
$r_{1/2}$ is the half-light radius determined in 2D) and dividing by the surface of the circle of this radius. This
scale is about the closest to the center of the dwarf possible, i.e. the position of the innermost data point in
measurements of surface brightness in real data. The results of this procedure are shown as solid lines in
Figure~\ref{sb1-7} for runs O1-O7.

We notice a similar degree of variability in the measured surface brightness values as was observed for the 2D
half-light radii. Since we measure the surface brightness within a fixed fraction of $r_{1/2}$ the two are obviously
tightly related. Note that in the case of simulations O2 and O4
the strong variability at the early stages is replaced by an
almost constant value between pericenters later on when the dwarf becomes spherical. The general trend is however to
decrease the central surface brightness which means that the stars are stripped from all radii, not only from the
outer parts i.e. the density profile of the stars is shifted down at all scales due to tidal stripping.

\subsection{Kinematics}

One of the key parameters that distinguish dIrrs from dSphs is their $V/\sigma$ value,
where $V$ is the rotation velocity and $\sigma$ is the (central) velocity dispersion of the stars.
While dIrrs are believed to be rotationally supported, with $V/\sigma$ of the order of a few, dSphs
usually exhibit much lower rotation levels with $V/\sigma$ below unity. However, contrary to common
belief, dSphs are not in general completely devoid of rotation and it has been detected for many
dSph galaxies, including Ursa Minor (Hargreaves et al. 1994; Armandroff et al. 1995),
Carina (Mu\~noz et al. 2006), Sculptor (Battaglia et al. 2008), Leo I (Sohn et al. 2007; Mateo et al. 2008),
Cetus (Lewis et al. 2007) and Tucana (Fraternali et al. 2009).

\begin{figure}
\begin{center}
    \leavevmode
    \epsfxsize=7.cm
    \epsfbox[50 10 230 180]{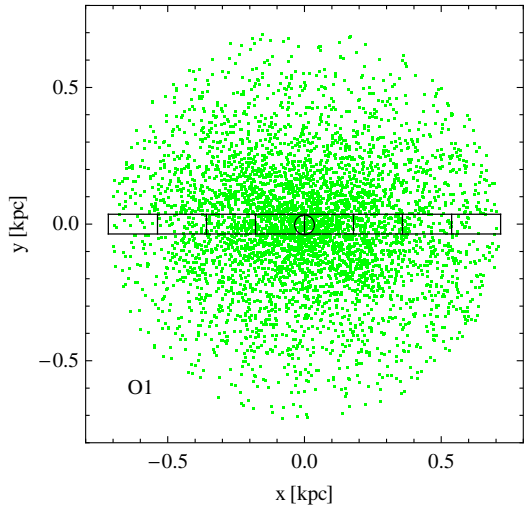}
\end{center}
\caption{The distribution of stars in simulation O1 within $2 r_{1/2}$ as seen from the galactic center
at the last apocenter of the orbit (green dots). The position of the stars were rotated so that the major axis of
the galaxy image is along the $x$ axis. Only 1 percent of the stars is shown for clarity. The inner black
circle of radius $0.1 r_{1/2}$ encloses stars that are used in the determination of the central velocity dispersion.
Black rectangles indicate eight bins along the major axis of the dwarf in which the rotation velocity and
the velocity dispersion profiles are measured.}
\label{kinematics}
\end{figure}

We measured $V$ and $\sigma$ using the following procedure, as close as possible to real observations.
For each simulation
output we selected stars, as seen by an observer at the galactic center, within projected radii $R < 2 r_{1/2}$
from the center of the dwarf, where $r_{1/2}$ are the 2D half-light radii determined above. This is approximately
the region where kinematic measurements for dwarf galaxies are typically performed. We then determined the
principal axes of the 2D distribution of the stars and rotated their positions in order to align the major axis
of the dwarf image with the horizontal ($x$) axis. Figure~\ref{kinematics} shows the distribution
of (1 percent) of the stars within $R < 2 r_{1/2}$ obtained in this way for the configuration corresponding to the
last apocenter for simulation O1.

\begin{figure}
\begin{center}
    \leavevmode
    \epsfxsize=8.6cm
    \epsfbox[20 10 290 686]{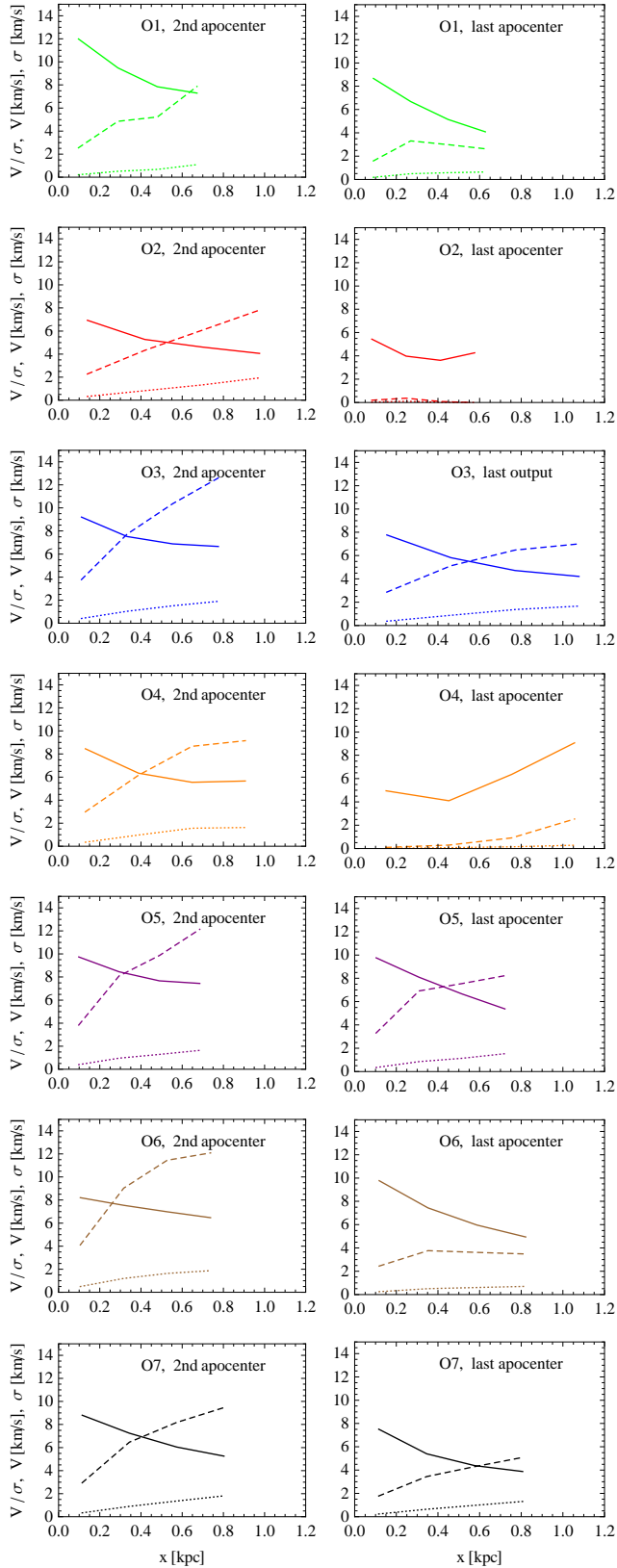}
\end{center}
\caption{The velocity dispersion profiles (solid lines), the rotation velocity profiles (dashed lines) and
the ratios of the two (dotted lines) measured along the major axis of the observed dwarf for
runs O1-O7 at their 2nd (left column) and last (right column) apocenters (except for the case O3 where the 2nd
apocenter is the last so we plot the result for the final simulation output).}
\label{vdprofiles1-7}
\end{figure}

\begin{figure*}
\begin{center}
    \leavevmode
    \epsfxsize=17cm
    \epsfbox[20 0 550 590]{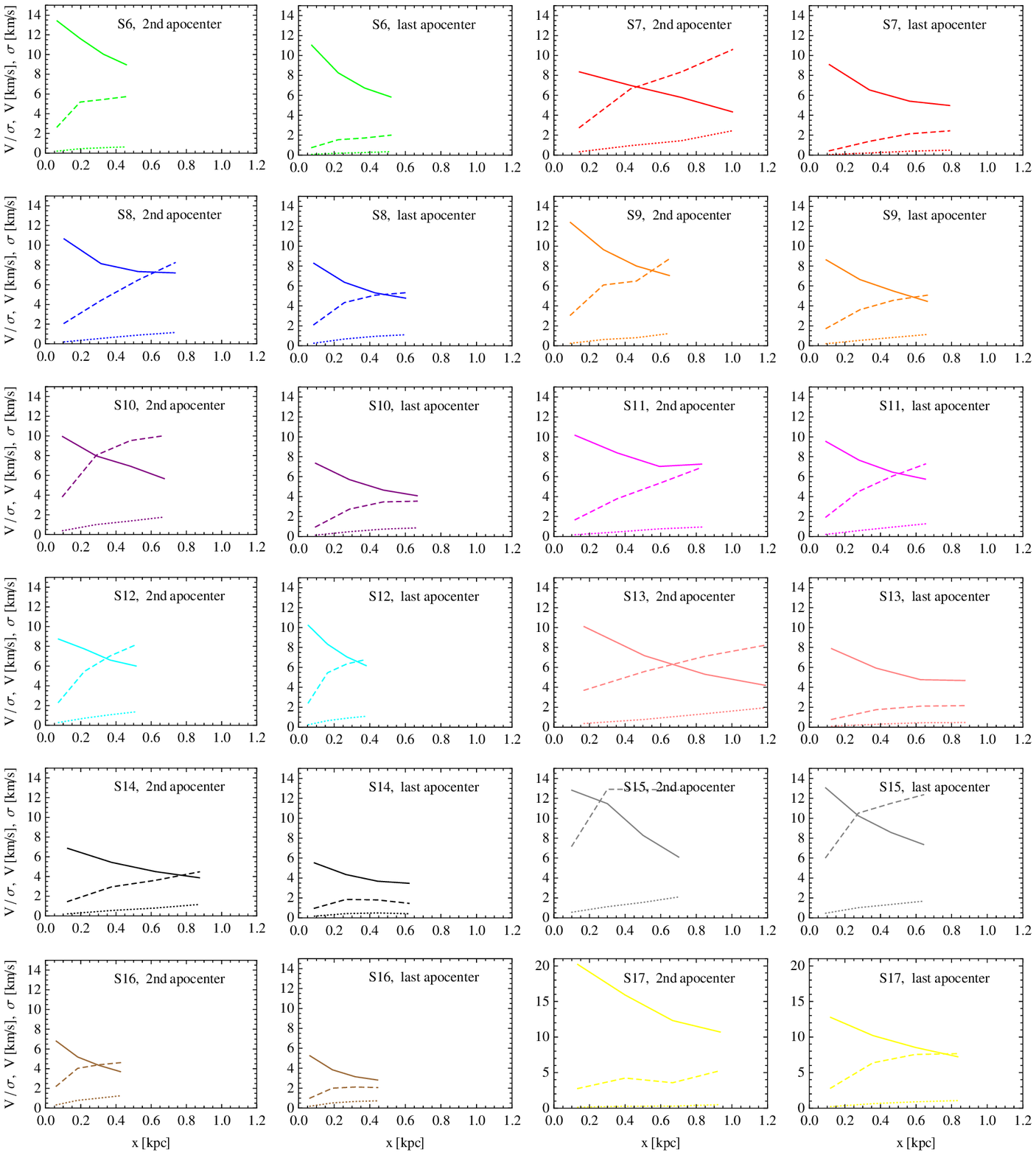}
\end{center}
\caption{The velocity dispersion profiles (solid lines), the rotation velocity profiles (dashed lines) and
the ratios of the two (dotted lines) measured along the major axis of the observed dwarf for
runs S6-S17 at their 2nd and last apocenters.}
\label{vdprofiless6-s17}
\end{figure*}

In order to estimate the central velocity dispersion of the dwarfs we selected for each output the stars within
$R < 0.1 r_{1/2}$ (small black circle in Figure~\ref{kinematics}) and calculated the
dispersion of radial velocity using standard estimators (see e.g. {\L}okas et al. 2005). By the
choice of this innermost region we make sure that the sample is not contaminated by tidally stripped stars
in the tails. Next, we select a narrow strip of stars along the major axis of the dwarf of size $0.1 r_{1/2}
\times 2 r_{1/2}$ and sample the velocities of the stars in eight bins of size $0.1 r_{1/2} \times 0.5 r_{1/2}$
each. Since the outer bins are likely to be contaminated by tidally stripped stars we introduce a cut-off in
velocities, i.e. we exclude the stars with velocities that differ from the dwarf's systemic velocity by more
than $3 \sigma$ where $\sigma$ is the central velocity dispersion measured in the inner $R < 0.1 r_{1/2}$.
Such a conservative cut-off is usually not sufficient for the reliable selection of stars for dynamical modelling
(see Klimentowski et al. 2007, 2009b; {\L}okas et al. 2008; {\L}okas 2009) but seems stringent enough to
measure basic kinematic properties of the dwarfs, as we aim to do here.

The results of these measurements for simulations O1-O7 and S6-S17 are shown in
Figures~\ref{vdprofiles1-7} and \ref{vdprofiless6-s17} respectively
for the outputs corresponding to the 2nd and the last apocenter (the left and right panel for each run).
With dashed lines we plot the rotation velocity profiles and with solid lines the velocity
dispersion profiles to illustrate their dependence on the
distance from the center of the dwarf. Dotted lines give the ratio of the two values as a function of radius.
All measurements were symmetrized, i.e. we show the average of two measurements
on both sides of the dwarf. In most cases, the rotation profile decreases significantly between the 2nd and the last
apocenter, while the dispersion remains roughly on the same level. This is due to the fact that tidal stirring
causes the rotation to be replaced by random motions increasing the velocity dispersion,
but at the same time dwarfs lose mass and
the dispersion is decreased. The net outcome of these two effects is that the dispersion remains roughly constant
in time.

All dispersion profiles look well-behaved, declining with radius,
except for the case O4 (the right panel of the fourth row in Figure~\ref{vdprofiles1-7}, orange line)
where the dispersion profile shows a secondary increase at larger distances from the center. This is due to the
contamination from the tidal tails that affects our measurements because the nominal half-light radius determined
for this case is very large. As already discussed above, and seen
in the fourth panel of Figure~\ref{radii1-7}, in this case the $r_{1/2}$ values
measured near apocenters at the later stages of evolution are about a factor of two larger than those at
pericenters. Again, this is caused by the presence of tidal tails that for this very radial orbit
($r_{\rm apo}/r_{\rm peri} = 10$, see Table~\ref{properties}) for most of the time (except at the very pericenter)
are oriented along the observer's line of sight. The dwarf thus appears brighter and more extended
(see also Figure~11 in Klimentowski et al. 2007). The contamination is also seen in the rotation curve
(dashed orange line in the same panel of Figure~\ref{vdprofiles1-7}) which shows tidally induced velocity
gradient in the outer bins while no rotation is detected in the inner part for this very evolved dwarf.

\begin{figure}
\begin{center}
    \leavevmode
    \epsfxsize=7.45cm
    \epsfbox[60 10 311 745]{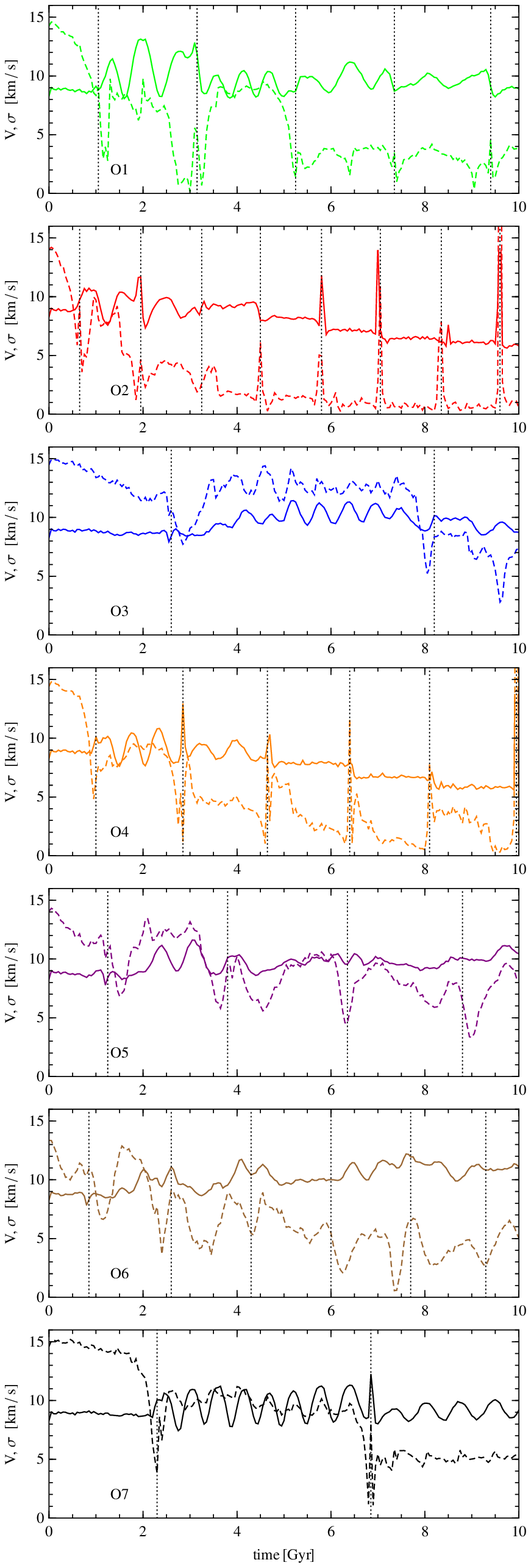}
\end{center}
\caption{The evolution of the central velocity dispersion (solid line) and the rotation velocity (dashed line)
in time for simulations O1-O7. The central velocity dispersion was measured for stars with projected radii
$R< 0.1 r_{1/2}$ and the rotation velocity was taken as a maximum value found along the major axis of the galaxy
image within $2 r_{1/2}$, as shown in the upper panel of Figure~\ref{kinematics}.
Vertical dotted lines indicate pericenter passages.}
\label{veldisp1-7}
\end{figure}

Figure~\ref{veldisp1-7} shows the evolution of the kinematic properties in time for the seven simulations O1-O7
with different orbits. In each panel the measurements of the central velocity dispersion are plotted with solid lines
while the values of the rotation velocity are shown with dashed lines. The values of rotation are taken as
a maximum rotation velocity found along the rotation velocity profile (see Figures~\ref{vdprofiles1-7}
and \ref{vdprofiless6-s17}) measured along the major axis within $R < 2 r_{1/2}$. In simulations O2 and O4
where the dwarfs are most strongly affected by tidal forces, the contamination by tidal tails is well
visible, although in a slightly different way in each run. For orbit O2 we see sharp increases of both velocity
dispersion and rotation velocity at later pericenters caused by spikes in the measured values of $r_{1/2}$.
In simulation O4, on the other hand, while the measured
values of the central velocity dispersion remain unaffected (until the very last pericenter), the rotation velocity
is overestimated when the dwarf is on its way from the pericenter to the
apocenter. As discussed above, this is due to the overestimated value of
$r_{1/2}$ which makes us probe the region well beyond the dwarf's main stellar body.

Finally, it is worth noting that the
tidal effects on the observables in runs O2 and O4, although in both cases due to tidally stripped stars,
are of slightly different origin; while in O2 they are due to earlier wraps of tidally lost material (this run
has more pericentric passages than any other orbit we considered), in O4 they are due to recently lost material
residing in the tails oriented radially because of the high eccentricity of the orbit. In other runs the most
characteristic feature of the kinematic measurements is the oscillation of the values (especially those of the central
velocity dispersion) in simulations due, as discussed above, to the non-sphericity of the stellar distribution and
the tumbling of its triaxial shape.

\begin{figure}
\begin{center}
    \leavevmode
    \epsfxsize=8.2cm
    \epsfbox[60 0 310 515]{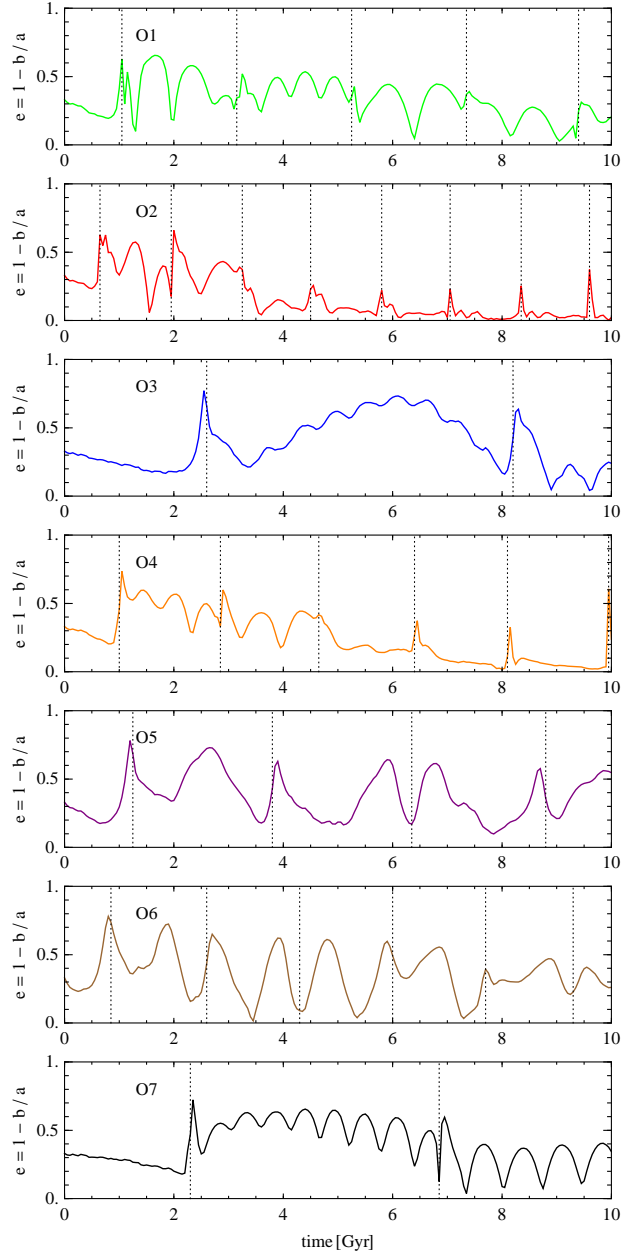}
\end{center}
\caption{The evolution of the ellipticity $e = 1-b/a$ in time for simulations O1-O7.
Measurements of the axis ratios $b/a$ are performed within $2 r_{1/2}$ where $r_{1/2}$ is determined by
an observer situated at the galactic center
within the angular aperture corresponding to the projected radius $R = r_{\rm lim}$.
Vertical dotted lines indicate pericenter passages.}
\label{ellipticity1-7}
\end{figure}

\subsection{Shapes}

We conclude the measurements of the observational parameters of our simulated dwarfs by determining their shapes.
For this purpose we select in each output the stars within projected radii $R < 2 r_{1/2}$ and rotate the 2D
distribution to align the major axis of the galaxy image with $x$ and the minor axis with $y$, exactly as for the
measurements of the kinematic properties. We then calculate the ellipticity parameter $e=1-b/a$ where $b/a$ is
the axis ratio of the 2D distribution determined from the eigenvalues of the 2D inertia tensor.

The evolution of the ellipticity parameter $e$ in time is shown in Figure~\ref{ellipticity1-7} for simulations
O1-O7. Note that all runs start from the same value of $e=0.33$ which is due to the specific orientation of the
dwarf galaxy disk at the beginning of the simulation. Namely, the disks are inclined by $45^\circ$ to the initial
orbital plane and the initial velocity vector so the galaxy images appear elliptical to the observer at the
center of the Milky Way. At later stages the ellipticity parameter displays strong variability due to the formation
of the triaxial shape and tumbling of this shape, as well as the variability of the $2 r_{1/2}$ scale inside which
we measure the shape. A clear transition to sphericity is however seen for runs O2 and O4 where the ellipticity
is very close to zero at the later stages of the evolution with momentary exceptions at pericenters where the scale
$2 r_{1/2}$ picks up some of the tidal tails oriented perpendicular to the line of sight at these instants.

\begin{figure}
\begin{center}
    \leavevmode
    \epsfxsize=8.2cm
    \epsfbox[60 0 310 220]{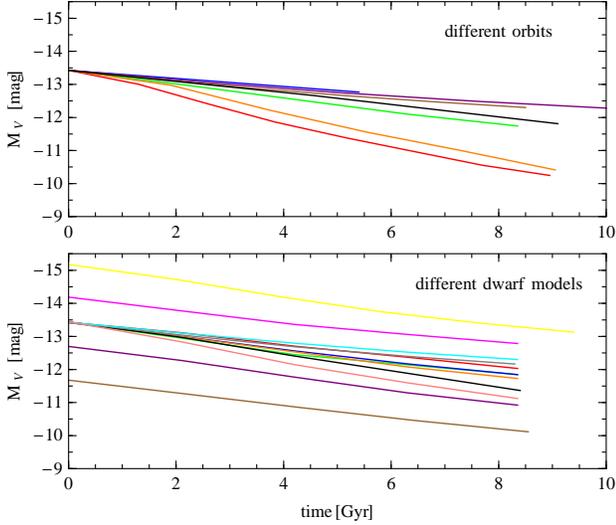}
\end{center}
\caption{The evolution of the total visual magnitude in time from the first to the last apocenter
for different orbits (simulations O1-O7, upper panel)
and different dwarf models (simulations S6-S17, lower panel).}
\label{magprojapo}
\end{figure}

\begin{figure}
\begin{center}
    \leavevmode
    \epsfxsize=8.2cm
    \epsfbox[60 0 310 220]{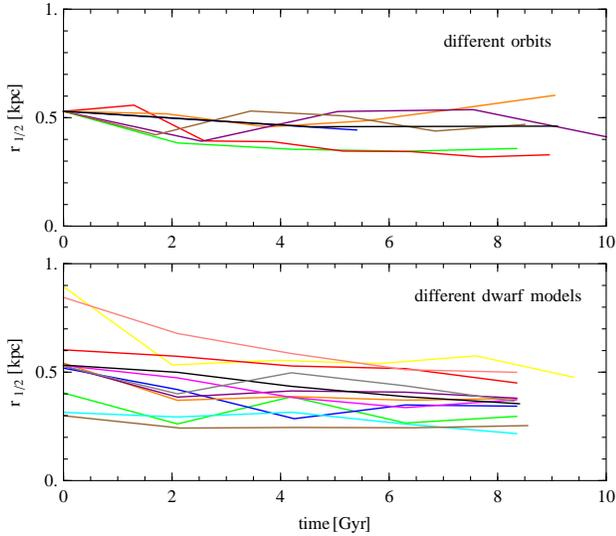}
\end{center}
\caption{The evolution of the half light radius in time from the first to the last apocenter
for different orbits (simulations O1-O7, upper panel)
and different dwarf models (simulations S6-S17, lower panel).}
\label{radiiprojapo}
\end{figure}

\begin{figure}
\begin{center}
    \leavevmode
    \epsfxsize=8.2cm
    \epsfbox[60 0 310 220]{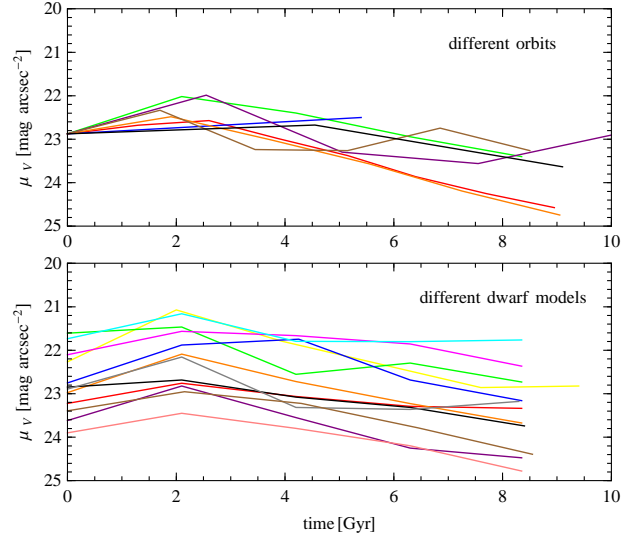}
\end{center}
\caption{The evolution of the central surface brightness in time from the first to the last apocenter
for different orbits (simulations O1-O7, upper panel)
and different dwarf models (simulations S6-S17, lower panel).}
\label{sbapo}
\end{figure}

\begin{figure}
\begin{center}
    \leavevmode
    \epsfxsize=8.cm
    \epsfbox[60 0 310 225]{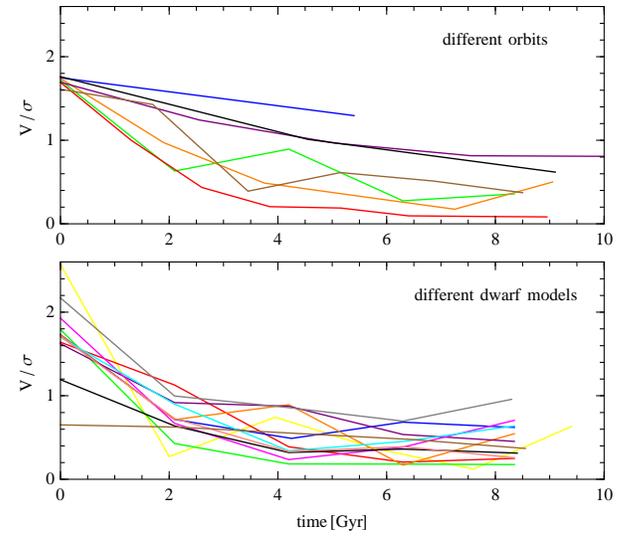}
\end{center}
\caption{The evolution of the ratio $V/\sigma$ in time from the first to the last apocenter
for different orbits (simulations O1-O7, upper panel)
and different dwarf models (simulations S6-S17, lower panel).}
\label{vsapo}
\end{figure}

\begin{figure}
\begin{center}
    \leavevmode
    \epsfxsize=8.2cm
    \epsfbox[60 0 310 220]{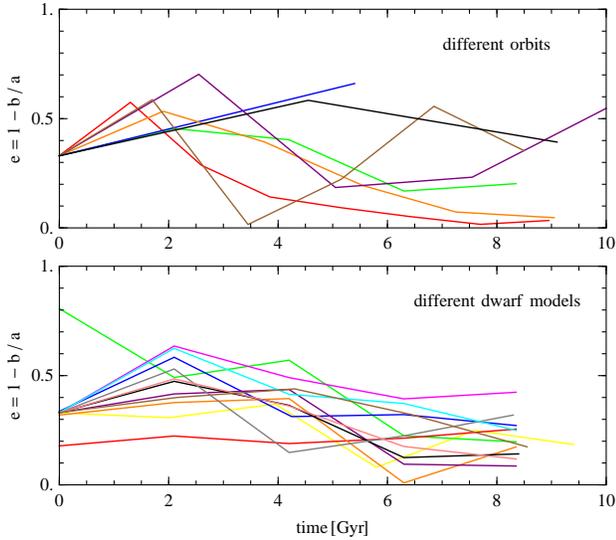}
\end{center}
\caption{The evolution of the ellipticity $e = 1-b/a$ in time from the first to the last apocenter
for different orbits (simulations O1-O7, upper panel)
and different dwarf models (simulations S6-S17, lower panel).}
\label{ellipticityapo}
\end{figure}

\section{Long-term evolution of the observed properties}

In order to study the general trends in the evolution of the observational parameters in time in
Figures~\ref{magprojapo}-\ref{ellipticityapo} we plot the values measured previously only at subsequent
apocenters, but for all simulations. In the upper panel of each figure we show with different colors the results for
the simulations for different orbits and the same initial dwarf structure O1-O7, while in the lower panel we plot
results for different initial structures of the dwarf and for the same default orbit S6-S17. A particular simulation
can be identified by referring to the colors listed in the last column of Table~\ref{properties}. The time of the last
apocenter for each simulation can be found in the sixth column of the Table and the corresponding 2D values of the
total magnitude $M_V$, half-light radius $r_{1/2}$, central surface brightness $\mu_V$, the ratio of the rotation
velocity to the central velocity dispersion $V/\sigma$ and ellipticity $e=1-b/a$ measured at this last
apocenter are listed in columns 8-12 of the Table.

The general trends in the evolution of the parameters are now clearly visible. As expected, the tidal
stripping decreases the total magnitude of the dwarf and most so for the tightest orbit O2 and the one with the
smallest pericenter and reasonably small orbital time O4. The trend of $r_{1/2}$ decreasing with time is less clear and
more visible for dwarfs of different structure rather than on different orbits. Overall, the half-light radius
remains remarkably constant over time, in contrast to the other characteristic scale-length of the dwarf, $r_{\rm
max}$, where the maximum of the circular velocity occurs which was found by K11 to decrease very strongly
during the evolution. This is not surprising, however, given that $r_{\rm max}$ is a characteristic scale-length
of the total gravitational potential, which is dominated by dark matter, while $r_{1/2}$ characterizes the distribution
of stars which are not so strongly affected by tidal forces.

The central surface brightness behaves differently from the other two properties discussed above. Typically, its value
increases between the first and the second apocenter and then starts to decrease. This is due to the formation of the
bar which in most cases occurs at the first pericenter and increases the density of the stars in the center of the dwarf.
Later on,
when the stellar distribution becomes more spherical and more stars are stripped, $\mu_V$ decreases as a function of time.

In Figure~\ref{vsapo} we plot the ratio of the rotation velocity and the central velocity dispersion, $V/\sigma$,
which measures the amount of ordered versus random motion of the stars. Both for different orbits and for different
dwarf models the trend of $V/\sigma$ decreasing with time is clearly visible. This is one of the key signatures of
tidal stirring and confirms that the process may indeed transform rotationally supported systems to those supported
by random motions.

Finally, Figure~\ref{ellipticityapo} shows the trends in the evolution of the observed ellipticity. Here the
trends are difficult to detect since a similar small ellipticity is measured for a disk close to face-on and a
genuine almost spherical galaxy. In almost all our simulations the disk was initially inclined so that the measured
ellipticity was $e=0.33$. The only exceptions are runs S6 where the disk is seen edge-on resulting in a large
initial ellipticity $e=0.81$ (green line in the lower panel of Figure~\ref{ellipticityapo}) and S7 where the disk
is more face-on than the default orientation resulting in a small initial $e=0.18$ (red line in the lower panel
of Figure~\ref{ellipticityapo}). For different orbits the transition to the almost perfectly spherical shape is
however reflected in very low $e$ values for runs O2 and O4, while for different dwarf models the ellipticity
averaged over all runs is smaller than initially.

\begin{table*}
\begin{center}
\caption{Properties of the dwarf galaxies of the Local Group. }
\begin{tabular}{llcccccl}
\hline
\hline
Dwarf	   & Type &  $M_V$ & $r_{1/2}$ & $\mu_V$             & $V/\sigma$ & $e=1-b/a$ & References  \\
galaxy     &      &  [mag] &  [kpc]    & [mag arcsec$^{-2}$] &            &           &             \\
\hline
WLM             &dIrr	    &$	-14.5 $ & 1.509 &  23.6	& 2.63	&	0.59   &  1,3         \\
IC 10           &dIrr	    &$	-15.7 $ & 0.775 &  22.1 & 3.75	&	0.3    &  1          \\
NGC 147         &dSph/dE    &$	-15.5 $ & 0.753 &  21.6	& 1.0	&	0.46   &  1,2        \\
NGC 185         &dSph/dE    &$	-15.5 $ & 0.521 &  20.1 & 0.63	&	0.26   &  1,2        \\
LGS 3           &dIrr/dSph  &$	-10.5 $ & 0.312 &  25.6	& 0.22	&	0.26   &  1,3        \\
IC 1613         &dIrr	    &$	-14.7 $ & 1.870 &  23.5	& 2.47	&	0.24   &  1,3        \\
Phoenix         &dIrr/dSph  &$	-10.1 $ & --    &  --   & 0.22  &      	0.3    &  1          \\
UGCA 92         &dIrr	    &$	-12.6 $ & 0.441 &  24.2 & 4.13  &      	0.55   &  1,4        \\
Leo A           &dIrr	    &$	-11.4 $ & 0.341 &  --   & 0.32  &      	0.40   &  1,19       \\
Sextans B       &dIrr	    &$	-14.2 $ & 0.417 &  22.9 & 1.22	&	0.23   &  1,3,4      \\
NGC 3109        &dIrr	    &$	-15.7 $ & 1.920 &  23.6 & 6.7   &      	0.80   &  1          \\
Antlia          &dIrr/dSph  &$	-10.8 $ & 0.672 &  24.3 & --    &      	0.35   &  1          \\
Sextans A       &dIrr	    &$	-14.6 $ & 1.068 &  23.7 & 2.38  &      	0.21   &  1,3        \\
GR 8            &dIrr	    &$	-11.6 $ & 0.189 &  22.3 & 0.64	&	0.31   &  1          \\
SagDIG          &dIrr	    &$	-12.3 $ & --    &  24.4	& 0.27	&	0.47   &  1          \\
NGC 6822        &dIrr	    &$	-15.2 $ & 0.582 &  21.4	& 5.88	&	0.47   &  1          \\
Aquarius        &dIrr/dSph  &$	-10.6 $ & 0.342 &  23.6	& 0.76	&	0.40   &  1,18        \\
IC 5152         &dIrr	    &$	-14.8 $ & --    &  --	& 3.88	&	0.18   &  1          \\
UGCA 438        &dIrr	    &$	-12.0 $ & 0.320 &  22.4	& --	&	0.05   &  1,4        \\
Pegasus         &dIrr/dSph  &$	-12.9 $ & --    &  23.7	& 1.16	&	0.40   &  1,3         \\
VV124           &dIrr/dSph  &$	-12.4 $ & 0.252 &  21.2 & 0.45  &       0.44   &  31          \\
Carina          &dSph	    &$	-8.62 $ & 0.241 &  25.5	& 0.43	&	0.33   &  1,5,6,7     \\
Draco           &dSph	    &$	-8.74 $ & 0.196 &  25.3 & 0.21	&	0.29   &  1,5,14,15   \\
Fornax          &dSph	    &$	-13.03$ & 0.668 &  23.4 & 0.18	&	0.31   &  1,5,7       \\
Leo I           &dSph	    &$	-11.49$ & 0.246 &  22.4 & 0.33  &       0.21   &  1,5,8      \\
Leo II          &dSph	    &$	-9.60 $ & 0.151 &  24.0	& 0.28	&	0.13   &  1,5,9       \\
Sculptor        &dSph	    &$	-10.53$ & 0.260 &  23.7 & 0.30	&	0.32   &  1,5,7,10    \\
Sextans         &dSph	    &$	-9.20 $ & 0.682 &  26.2 & 0.48	&	0.35   &  1,5,7,11    \\
Ursa Minor      &dSph	    &$	-8.42 $ & 0.280 &  25.5	& 0.49	&	0.56   &  1,5,12,13   \\
Sagittarius     &dSph	    &$	-13.3 $ & 1.550 &  25.2 & 0.50	&	0.63   &  5,16,17     \\
Andromeda I     &dSph	    &$	-11.8 $ & 0.600 &  24.7	& --	&	0.22   &  20         \\
Andromeda II    &dSph	    &$	-12.6 $ & 1.060 &  24.5	& --	&	0.20   &  20         \\
Andromeda III   &dSph	    &$	-10.2 $ & 0.360 &  24.8	& --	&	0.52   &  20         \\
Andromeda V     &dSph	    &$	-9.6  $ & 0.300 &  25.3	& --	&	0.18   &  20         \\
Andromeda VI    &dSph	    &$	-11.5 $ & 0.420 &  24.1	& --	&	0.41   &  20         \\
Andromeda VII   &dSph	    &$	-13.3 $ & 0.740 &  23.2	& --	&	0.13   &  20         \\
Cetus           &dSph	    &$	-11.3 $ & 0.600 &  25.0	& 0.45	&	0.33   &  20,22      \\
Tucana          &dSph	    &$	-9.5  $ & 0.274 &  25.1 & 1.04  &       0.48   &  1,21,23     \\
Andromeda IX    &dSph	    &$	-8.3  $ & 0.530 &  26.8	& --	&	0.25   &  21,25       \\
Andromeda X     &dSph	    &$	-8.1  $ & 0.339 &  26.7	& --	&	0.44   &  21,25,26    \\
Andromeda XIV   &dSph	    &$	-8.3  $ & 0.413 &  --   & --    &       0.31   &  21,27       \\
Andromeda XV    &dSph	    &$	-9.4  $ & 0.220 &  --	& --	&	--     &  21          \\
Andromeda XVI   &dSph	    &$	-9.2  $ & 0.136 &  --	& --	&	--     &  21          \\
Andromeda XVII  &dSph	    &$	-8.5  $ & 0.254 &  26.1	& --	&	0.27   &  21,26,30    \\
Andromeda XVIII &dSph	    &$	-9.7  $ & 0.363 &  25.6	& --	&	--     &  29          \\
Andromeda XIX   &dSph	    &$	-9.3  $ & 2.065 &  29.3	& --	&	0.17   &  29          \\
Andromeda XXI   &dSph	    &$	-9.9  $ & 0.875 &  27.0	& --	&	0.20   &  28          \\
Andromeda XXIII &dSph	    &$	-10.2 $ & 1.035 &  28.0	& --	&	0.40   &  24	    \\
Andromeda XXV   &dSph	    &$	-9.7  $ & 0.732 &  27.1	& --	&	0.25   &  24	    \\
CVnI            &dSph	    &$	-8.6  $ & 0.564 &  28.0	& --	&	0.38   &  21,32       \\
Leo T           &dIrr/dSph  &$	-8.1  $ & 0.115 &  26.9	& --	&	0.1    &  21,33,34    \\
\hline
\label{realdwarfs}
\end{tabular}
\end{center}
\tablerefs{
(1) Mateo 1998;
(2) Geha et al. 2010;
(3) Hunter et al. 2010;
(4) Sharina et al. 2008;
(5) Walker et al. 2010;
(6) Mu\~noz et al. 2006;
(7) {\L}okas 2009;
(8) {\L}okas et al. 2008;
(9) Koch et al. 2007;
(10) Battaglia et al. 2008;
(11) Battaglia et al. 2011;
(12) Mu\~noz et al. 2005;
(13) Wilkinson et al. 2004;
(14) {\L}okas et al. 2005;
(15) Kleyna et al. 2002;
(16) Majewski et al. 2003;
(17) {\L}okas et al. 2010b;
(18) McConnachie et al. 2006;
(19) Vansevi\v{c}ius et al. 2004;
(20) McConnachie \& Irwin 2006;
(21) Kalirai et al. 2010;
(22) Lewis et al. 2007;
(23) Fraternali et al. 2009;
(24) Richardson et al. 2011;
(25) Zucker et al. 2007;
(26) Brasseur et al. 2011;
(27) Majewski et al. 2007;
(28) Martin et al. 2009;
(29) McConnachie et al. 2008;
(30) Irwin et al. 2008;
(31) Bellazzini et al. 2011;
(32) Zucker et al. 2006;
(33) Irwin et al. 2007;
(34) de Jong et al. 2008.
}
\end{table*}

\section{Comparison with observations}

\begin{figure}
\begin{center}
    \leavevmode
    \epsfxsize=7.cm
    \epsfbox[10 10 210 420]{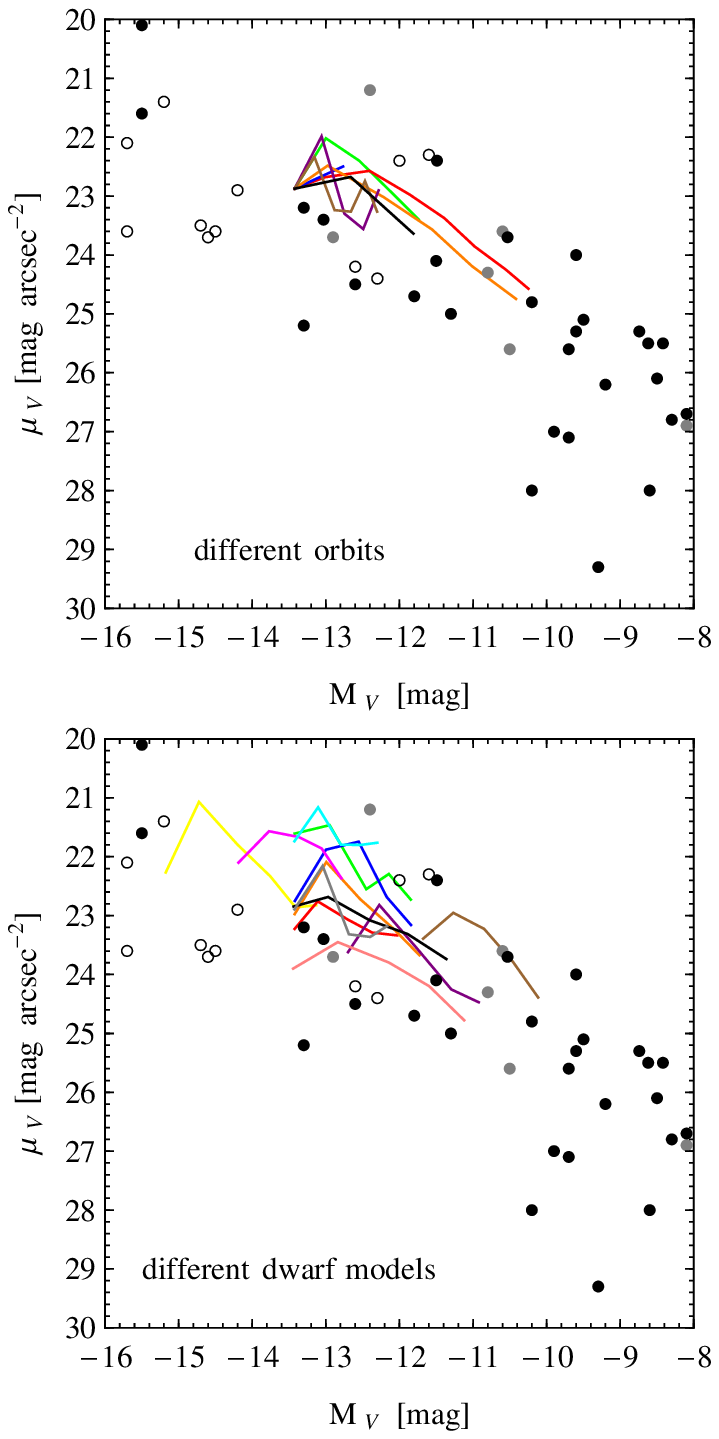}
\end{center}
\caption{The central surface brightness as a function of the total magnitude. Colored lines show the
results for the simulated dwarfs at subsequent apocenters for different orbits (simulations O1-O7, upper panel)
and different dwarf models (simulations S6-S17, lower panel). The evolution proceeds from the left to the right, from
the brighter to the fainter magnitudes in all cases.
In both panels the open circles show the data for dIrr galaxies,
filled black circles the data for the dSph and dSph/dE galaxies and gray circles for transitory dIrr/dSph dwarfs.
All data for the real dwarfs are listed in Table~\ref{realdwarfs}.}
\label{sbmagapo}
\end{figure}

\begin{figure}
\begin{center}
    \leavevmode
    \epsfxsize=7.cm
    \epsfbox[10 10 210 420]{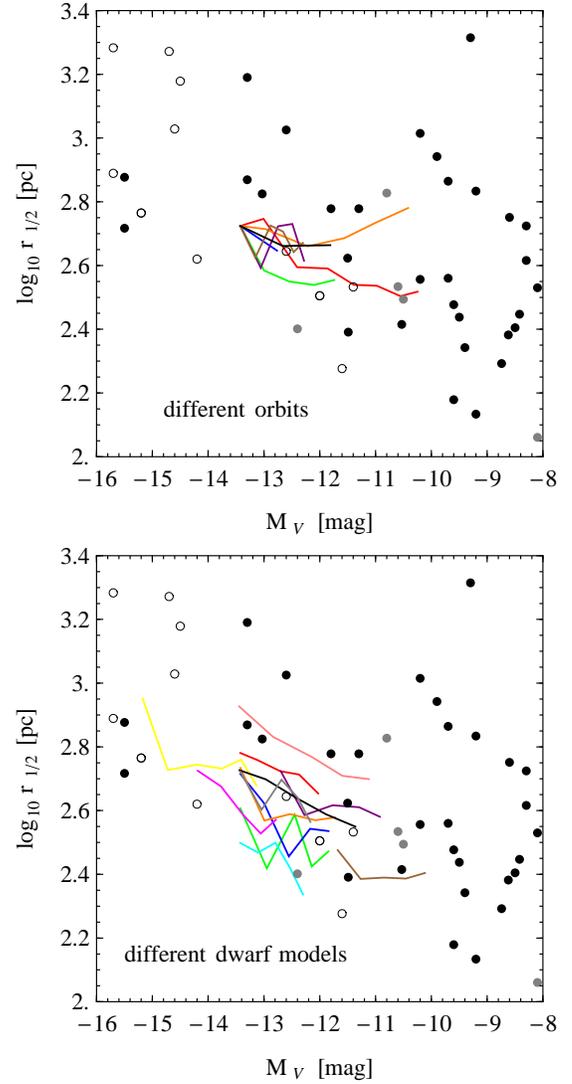}
\end{center}
\caption{Same as Figure~\ref{sbmagapo} but for the half-light radius as a function of the total
magnitude.}
\label{r12magapo}
\end{figure}

\begin{figure}
\begin{center}
    \leavevmode
    \epsfxsize=7.cm
    \epsfbox[10 10 210 420]{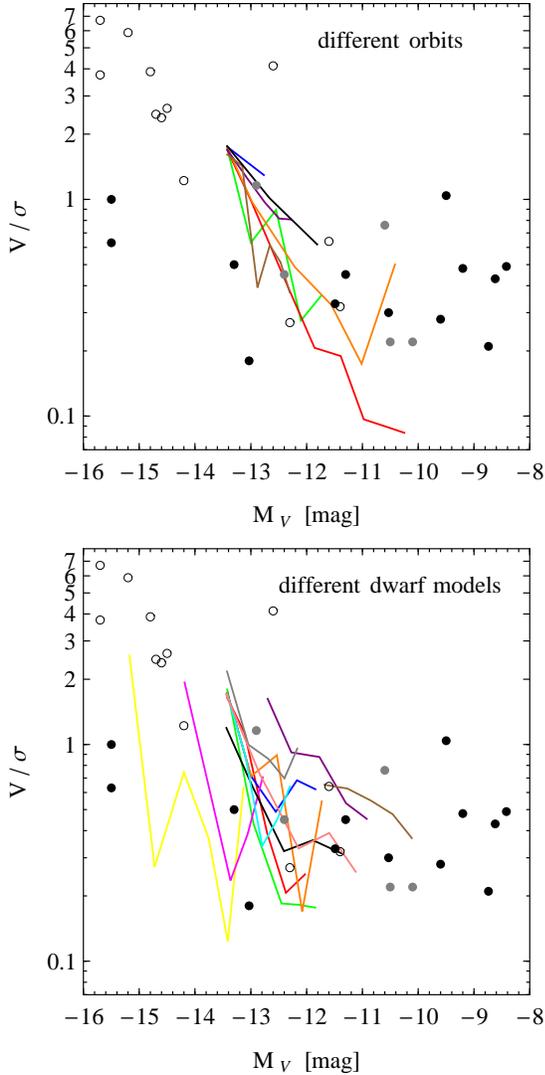}
\end{center}
\caption{Same as Figure~\ref{sbmagapo} but for the kinematic parameter $V/\sigma$ as a function of the total
magnitude.}
\label{vsmagapo}
\end{figure}

\begin{figure}
\begin{center}
    \leavevmode
    \epsfxsize=7.cm
    \epsfbox[10 10 210 420]{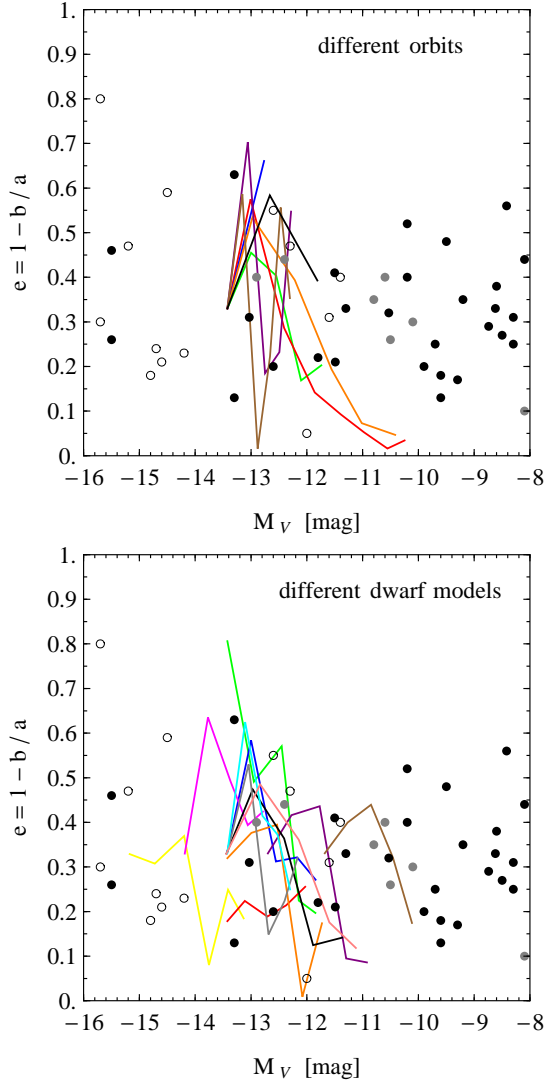}
\end{center}
\caption{Same as Figure~\ref{sbmagapo} but for the ellipticity parameter $e=1-b/a$ as a function of the total
magnitude.}
\label{ellimagapo}
\end{figure}

The ultimate purpose of our study is to compare the measured observational parameters of the simulated dwarfs
to real data.
In Figures~\ref{sbmagapo}-\ref{ellimagapo} we plot the central surface brightness $\mu_V$, the 2D half-light
radius $r_{1/2}$, the kinematic parameter $V/\sigma$ and the ellipticity $e=1-b/a$
respectively as a function of the total magnitude of the simulated dwarfs at subsequent apocenters. The color
coding of the simulations is the same as in previous Figures (see the last column of Table~\ref{properties})
and again the
upper panel is for dwarfs on different orbits while the lower one for dwarfs of different structure on the same default
orbit. The direction of the lines from the left to the right, i.e. from the brighter to the fainter magnitudes,
corresponds to the time flow because the total magnitude becomes fainter with time in all cases.

In Table ~\ref{realdwarfs} we list the properties of dwarf galaxies of the Local Group with magnitudes in the range
$-16 < M_V < -8$ currently available in the literature. The columns give the galaxy name, the morphological type,
the total visual magnitude, the half-light radius, the central surface brightness, the kinematic parameter
$V/\sigma$ and the ellipticity parameter $e=1-b/a$. The references are provided in the last column of the Table.
In cases where the exponential scale-lengths were only available, they were multiplied by a factor of 1.7 to
approximate the half-light radius of the exponential profile. For the $V/\sigma$, the `raw' values are used, i.e.
those measured directly from the data, as we do for our simulated dwarfs, without correcting for inclination.

\begin{figure*}
\begin{center}
    \leavevmode
    \epsfxsize=15.cm
    \epsfbox[40 0 410 335]{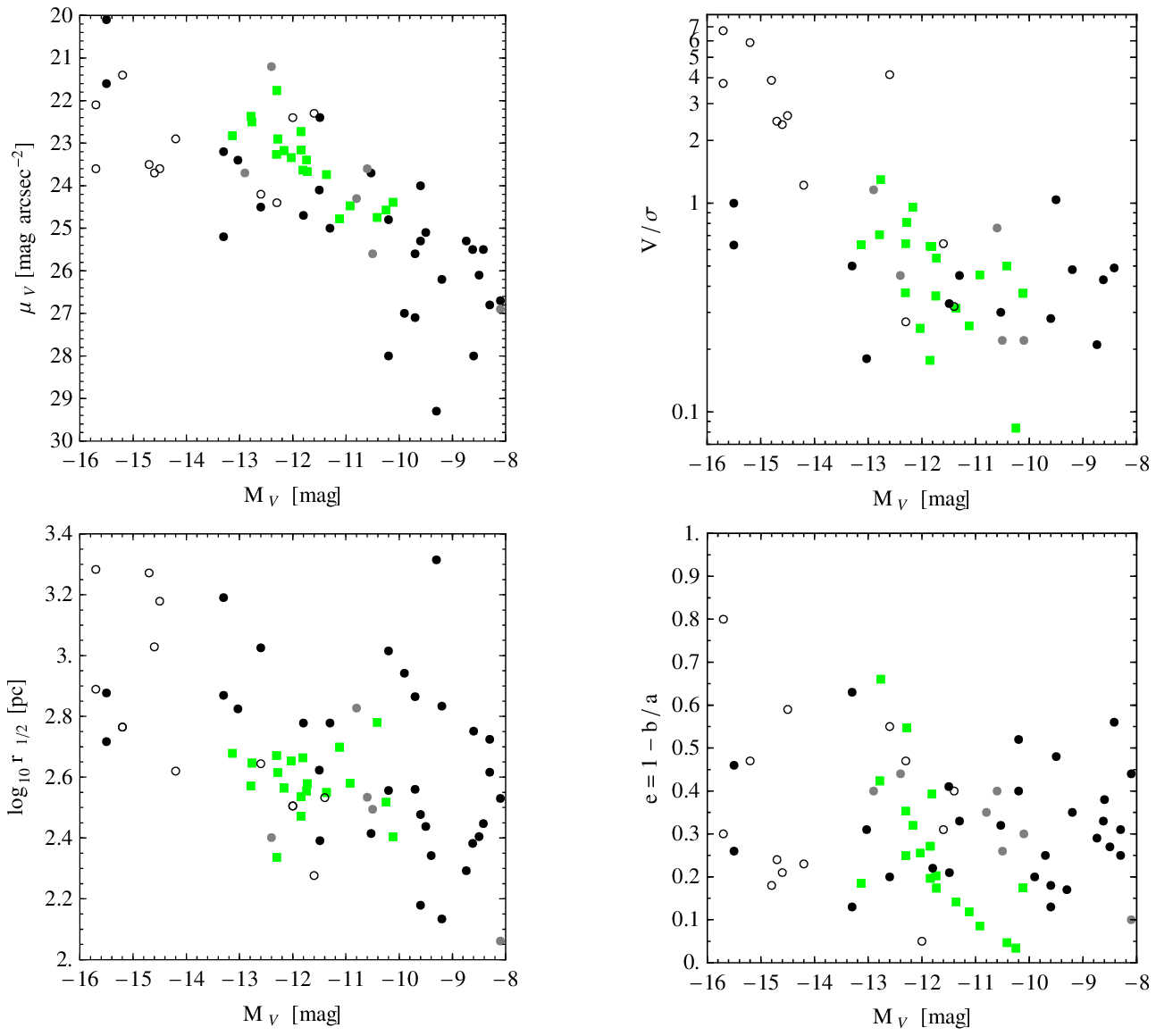}
\end{center}
\caption{The central surface brightness (upper left panel), the half-light radius (lower left panel),
the $V/\sigma$ parameter (upper right panel) and the ellipticity (lower right panel)
as a function of the total magnitude. Green squares show the measured parameters of the final products of tidal
stirring in our simulations (at the last apocenter of their orbit). The open black circles show the data for dIrr galaxies,
filled black circles the data for the dSph and dSph/dE galaxies and gray circles for transitory dIrr/dSph dwarfs
from Table~\ref{realdwarfs}.}
\label{allmag}
\end{figure*}

The circles in Figures~\ref{sbmagapo}-\ref{ellimagapo} show the real data for dwarf galaxies in the Local Group
from Table~\ref{realdwarfs}. The open circles mark the data for dIrr galaxies,
filled black circles for the dSph and dSph/dE galaxies and gray circles for transitory dIrr/dSph dwarfs.
The data in the $\mu_V$-$M_V$ and $r_{1/2}$-$M_V$ plots show strong correlation between the parameters (although
in the latter case the scatter is much larger) demonstrating the trends of fainter galaxies to possess lower
surface brightness and smaller size. There is also an obvious trend in $V/\sigma$-$M_V$: rotationally supported systems
are typically brighter. The correlation is weakest in the $e$-$M_V$ plane due to the difficulty in relating the
observed ellipticity to the real shape of a galaxy, as discussed above.

The evolutionary tracks of the simulated dwarfs in the $\mu_V$-$M_V$ plane (Figure~\ref{sbmagapo}) have a characteristic
shape, similar for all simulations. While the magnitudes evolve towards the fainter end monotonically, the values of
the surface brightness between the first and second apocenter increase and then mildly decrease at all subsequent
apocenters. This later evolution traces very well the correlation between the observed values of the parameters for
real Local Group dwarfs and is especially well visible in the case of runs O2 and O4 (the red and orange line in
Figure~\ref{sbmagapo}) where the dwarfs are most affected by tides and therefore most evolved.

In the $r_{1/2}$-$M_V$ plane (Figure~\ref{r12magapo}) the evolution of the simulated dwarfs shows more variety.
Typically, the characteristic radii decrease strongly between the first and second pericenter, but in the later
evolution they can also increase. Interestingly, because of the contamination by tidal tails,
it can also happen for the most affected dwarfs, like the one of run
O4 (the orange line in the upper panel of Figure~\ref{r12magapo}), where $r_{1/2}$ increases during the later
evolution, in contrast to the case of O2 (the red line). Overall, different orbits and different initial structures of
the dwarf are able to reproduce the trends seen in the real data.

The direction of the
evolution of the simulated dwarfs in the $r_{1/2}$-$M_V$ and especially in the
$\mu_V$-$M_V$ plane strongly suggests that the observed correlations may be
easily explained if dSph galaxies of the Local Group evolved from late-type dwarfs as predicted by the tidal stirring
scenario. The trend of surface brightness decreasing with decreasing luminosity as observed in
real dSph galaxies, in contrast to ellipticals, has been
interpreted as pointing towards different formation scenarios of spheroidal and elliptical galaxies by Kormendy (1985)
and Kormendy et al. (2009). They suggested that elliptical galaxies may form mostly via mergers while spheroidals are
rather late-type systems that underwent transformation due to star-formation processes or environmental effects such as
tidal stirring. This idea is supported by the analysis of the simulated
Local Group by Klimentowski et al. (2010) who found that mergers of subhalos are rather rare in such
systems and occur early
on so they cannot significantly contribute to the formation of a large fraction of dSph galaxies. The tidal
stirring scenario thus seems to be the most effective gravitational mechanism by which such objects could form.

\section{Discussion}

\subsection{Observed versus intrinsic properties}

Our results demonstrate the difficulty in determining the intrinsic properties of tidally stirred dwarf galaxies
from the photometric and kinematic measurements alone. The interpretation of the observational biases in the
photometric properties such as total magnitudes, half-light radii and central surface brightness is straightforward.
All of them are affected by three issues: the presence of contamination by tidal tails, the non-sphericity of the
stellar distribution and the `proximity effect' that for strongly evolved dwarfs on orbits with small pericenters
makes the observer seeing the dwarf at pericenter include more stars than are actually associated with the dwarf.

The measurements of kinematics and shapes are more complicated to interpret.
Although in both these quantities the trend towards
smaller $V/\sigma$ (signifying less rotation and more random motion in the stars) and towards more spherical shapes is
clearly seen especially for more evolved dwarfs, these measurements depend critically on the initial orientation
of the dwarf disks. This orientation was the same for all our simulations except S6 and S7 and may be considered
`typical' in the sense that our observer at the galactic center is able to detect both rotation and non-spherical shape
of the stellar component at the initial state and later.

In particular, the cases O3, O5 and S15 classified as `non-dSph' by K11 (see cases R3, R5 and R15 in their Table~2)
based on their high {\em intrinsic\/} $V/\sigma>1$ (measured as the mean rotation around the shortest axis
and the mean velocity dispersion for stars inside $r_{\rm max}$)
also show high values of the order of unity in our measurements so they would be
classified as transitory objects also by our observer.
However, if the disk was seen exactly edge-on (as in run S6) the observer
could identify it as rotating disk based on shape and rotation alone, while if it was face-on our observer
would not be able to distinguish the disk from a spheroid supported by random motions. In real observations such
cases can be morphologically classified by looking for additional signatures like the presence of gas or star
formation regions.

Despite these difficulties, we emphasize that in this study we measured the properties of the simulated
dwarfs in a realistic way so that they can be directly compared to real data. Figure~\ref{allmag} summarizes
our results by showing
as green squares the measured properties of the simulated dwarfs at the last apocenter in the
$\mu_V$-$M_V$, $r_{1/2}$-$M_V$, $V/\sigma$-$M_V$ and $e$-$M_V$ planes. The circles display
the data for dwarf galaxies of the Local Group from Table~\ref{realdwarfs} as in
Figures~\ref{sbmagapo}-\ref{ellimagapo}. The comparison proves that our dwarfs are indeed very similar to the real
ones in those crucial parameters and thus shows that tidal stirring indeed can produce dSph galaxies from dIrr-like
progenitors. Note that all our dwarfs (except two) initially had rather high masses of $10^9$ M$_{\odot}$ and therefore
ended up on the more luminous side of the observed distribution. Starting from lower masses and luminosities
we could easily reproduce also the less luminous dwarfs.

\subsection{Comparison with earlier work}

Observational parameters of the dSph galaxies produced by tidal stirring have already been discussed in the
papers proposing the tidal stirring as a possible scenario for the formation of dSph galaxies by Mayer
et al. (2001a,b). They have shown that the properties of the final products indeed follow the $\mu_V-M_V$ relation
found for real dwarfs and that the $V/\sigma$ and ellipticity both decrease with time. {\L}okas et al. (2010a)
measured the observational properties of dSph galaxies formed in three simulations discussed in Klimentowski
et al. (2009a), similar to those studied here, and found them to depend strongly on the line of sight
but still in the range characteristic of real dwarfs. In this work we improved on these
earlier results by showing the actual evolutionary tracks of the dwarfs in different parameter planes and considering
different observational biases caused by the observer's position close to the galactic center.

Pe\~narrubia et al. (2008) studied the effect of tides on objects that were spherical from the start and
constructed as stellar King profiles embedded in more extended dark halos. Understandably, their dwarfs do not
undergo any morphological transformation and remain spherical and supported by random motions with no net rotation.
Thus, they were unable to address such issues as the evolution of rotation or ellipticity. Their analysis was
therefore restricted to such observational parameters as the total luminosity, the half-light radius, surface
brightness and velocity dispersion. The evolutionary tracks of their spherical dwarfs in the $\mu_V$-$M_V$ and
$r_{1/2}$-$M_V$ planes (their Figure 10) are similar as in our simulations (Figures~\ref{sbmagapo}-\ref{r12magapo}),
i.e. in both cases there is a clear trend of surface brightness and radii decreasing with decreasing luminosity
of the dwarfs in time. Interestingly, the characteristic scales of the stellar distribution, quantified by the core
radius of the King profile, were found there to evolve rather mildly and decrease only by about a factor of two,
as in the case of our half-light radius.

Despite these similarities in the photometric properties, we find very different results concerning the
evolution of the velocity dispersion and the mass-to-light ($M/L$) ratio. Since we were more interested in the evolution
of the kinematic parameter $V/\sigma$ we do not show analogous plots for the dispersion alone, but the corresponding
trends can be read from Figure~\ref{veldisp1-7} where both quantities, $V$ and $\sigma$, are plotted separately.
Clearly, the central velocity dispersion decreases systematically only for the most strongly evolved dwarfs
(O2 and O4), while it remains roughly
constant for the other cases, in contrast to the findings of Pe\~narrubia et al. The evolution of the $M/L$ ratio
as an intrinsic, rather than observational, parameter of our dwarfs was discussed already by K11: they found that
tidal stripping typically decreases $M/L$ except for strongly evolved dwarfs where the dark halo has been truncated
down to the stellar scales. Again, our results on this point are discrepant with those of Pe\~narrubia et al. who
found $M/L$ increasing, making the dwarfs darker with time. The differences in the evolution
are due to different initial conditions: in our case the stars are more tightly bound (especially
after the formation of the bar) and are therefore more difficult
to strip, keeping the velocity dispersion roughly constant and $M/L$ rather low.
In the King models of Pe\~narrubia et al. the stars
are weakly bound and easily stripped which decreases their dispersion and increases $M/L$.
In summary, while the evolution of the photometric
properties seems to be a general feature of tidal stripping, largely independent of the initial structure of the dwarfs,
the kinematic and dynamical properties depend on the assumed dwarf model.

\subsection{Conclusions}

We have studied the observational parameters of simulated tidally stirred dwarf galaxies initially composed of disks
embedded in extended dark matter halos. Our results build on and extend the previous study by K11, where
intrinsic properties of the simulated dwarfs were discussed. Our main conclusions may be summarized as follows:
\begin{enumerate}

\item
The photometric parameters, such as the absolute magnitude, the half-light radius and the central surface brightness
as well as the kinematic properties
can be reliably measured but are subject to a number of observational biases caused by the presence of tidally
stripped stars and the non-sphericity of the stellar component.

\item
The amount of bias in the measured quantities depends on the orbit and structural parameters of the dwarf but
typically does not exceed 0.4 mag in total visual magnitude and a factor of two in half-light radius. Rotation can
be overestimated for strongly stripped dwarfs on eccentric orbits even if a $3\sigma$ cut-off is applied to radial
velocities.

\item
The effects of tidal stirring manifest themselves by decreasing the observed total magnitude, the half-light radius,
the central surface brightness, rotation and ellipticity in time, while the central velocity dispersion remains roughly
constant except for the most heavily stripped dwarfs where it also decreases due to strong mass loss.

\item
The correlations between the observational parameters of the simulated dwarfs reveal strong similarity
to those in the real data. The evolutionary tracks of the tidally stirred dwarfs move them from regions characteristic
for dIrr galaxies to those typical for dSph galaxies of the Local Group. This behavior corroborates the existence of a
connection between these two types of objects and therefore strongly supports the tidal stirring model as
a possible scenario of their evolution.

\end{enumerate}

\acknowledgments

This research was partially supported by the Polish National Science Centre under grant N N203 580940
and the Polish Ministry of Science and Higher Education under grant 92/N-–ASTROSIM/2008/0.
S.K. is funded by the Center for Cosmology and Astro-Particle Physics (CCAPP) at The Ohio State University.
The numerical simulations were performed on the Cosmos cluster at the Jet Propulsion Laboratory (JPL).
This work also benefited from an allocation of computing time from the Ohio Supercomputer
Center (http://www.osc.edu).

\end{document}